\documentclass[10pt,journal,compsoc]{IEEEtran}
\usepackage[normalem]{ulem}
\usepackage{pdfpages}

\ifCLASSOPTIONcompsoc
  \usepackage[nocompress]{cite}
\else
  \usepackage{cite}
\fi
\ifCLASSINFOpdf
\else
\fi

\usepackage{hyperref}

\usepackage{ragged2e} 
\usepackage[utf8]{inputenc} 

\usepackage{float}
\usepackage{graphicx}
\usepackage[font=footnotesize,labelfont=bf]{caption}
\usepackage{color,soul} 

\usepackage{textcomp}
\usepackage{multirow, makecell}
\usepackage{rotating}
\usepackage{adjustbox}
\usepackage{subcaption}

\usepackage[figurename=Figure]{caption}
\usepackage[tablename=Table]{caption}
\usepackage{url}
\usepackage[utf8]{inputenc}

\usepackage{enumitem}
\setlist[itemize]{leftmargin=*}

\begin{document}

\bstctlcite{IEEEexample:BSTcontrol}

%
\title{GateKeeper-GPU: Fast and Accurate Pre-Alignment Filtering in Short Read Mapping}
%
%
%
%

\author{Zülal~Bingöl, 
        Mohammed Alser, 
        Onur Mutlu, 
        Ozcan Ozturk, 
        and~Can~Alkan

\IEEEcompsocitemizethanks{
\IEEEcompsocthanksitem Z. Bingöl and C. Alkan are with the Department of Computer Engineering, Bilkent University, Ankara, Turkey. Z.Bingöl's email: zulal.bingol@bilkent.edu.tr, C.Alkan's email: calkan@cs.bilkent.edu.tr
\IEEEcompsocthanksitem O. Ozturk is with the Computer Science and Engineering and Electronics Engineering programs at Sabancı University, Istanbul, Turkey, and with the Department of Computer Engineering, Bilkent University, Ankara, Turkey. Email: ozturk@cs.bilkent.edu.tr
\IEEEcompsocthanksitem M. Alser and O. Mutlu are with the Department of Information Technology and Electrical Engineering, ETH Zürich, Switzerland. M.Alser's email: alserm@ethz.ch, O.Mutlu's email: onur.mutlu@safari.ethz.ch}

}

\IEEEtitleabstractindextext{%

\begin{abstract}
\justifying
At the last step of short read mapping, the candidate locations of the reads on the reference genome are verified to compute their differences from the corresponding reference segments using sequence alignment algorithms. Calculating the similarities and differences between two sequences is still computationally expensive since approximate string matching techniques traditionally inherit dynamic programming algorithms with quadratic time and space complexity. We introduce GateKeeper-GPU, a fast and accurate pre-alignment filter that efficiently reduces the need for expensive sequence alignment. GateKeeper-GPU provides two main contributions: first, improving the filtering accuracy of GateKeeper~(a lightweight pre-alignment filter), and second, exploiting the massive parallelism provided by the large number of GPU threads of modern GPUs to examine numerous sequence pairs rapidly and concurrently. By reducing the work, GateKeeper-GPU provides an acceleration of 2.9× to sequence alignment and up to $1.4\times$ speedup to the end-to-end execution time of a comprehensive read mapper~(mrFAST). GateKeeper-GPU is available at https://github.com/BilkentCompGen/GateKeeper-GPU
\end{abstract}

}

\maketitle

\IEEEdisplaynontitleabstractindextext

%
\IEEEpeerreviewmaketitle

\IEEEraisesectionheading{\section{Introduction}\label{sec:introduction}}

%
%
%
%
 
\IEEEPARstart{H}{igh-}throughput sequencing~(HTS) is the standard for deep and detailed studies in many areas such as population genomics and precision medicine. As the vision of predictive, preventive, personalized, and participatory~(P4) medicine~\cite{flores2013p4} rapidly becomes more prevalent, sequencing technologies stand out to provide vital information as a base for further investigation. Sequencing data generated by HTS constitutes a reliable source for biological research pipelines. 

As a result of massively parallel sequencing, a tremendous amount of data can now be generated in a single run~\cite{xin2013accelerating, arram2017leveraging}, making genomics one of the largest sources of big data today and in the future~\cite{stephens2015big}. On the other hand, the enormous data creates a computational challenge for analysis regarding both runtime and memory footprint, emphasizing the importance of developing efficient methods for quickly and efficiently analyzing genomic data.

The first main step in genome analysis is either \textit{de novo assembly}~\cite{chaisson2015genetic} or \textit{read mapping}, depending on the goal of the study. For read mapping, the reference genome of the subject species stands as a template~\cite{ruffalo2011comparative}, and newly sequenced fragments of a genome are compared to it to understand the genetic material of the individual sample. Read mapping consists of two main stages: \textit{seeding} and \textit{verification} (i.e., sequence or read alignment). In the seeding stage, the potential locations of reads on the reference genome are found based on the string similarity between the reads and corresponding reference segments~\cite{alser2021technology}. Reads generally have more than one possible seed location on the reference genome due to genomic repeats~\cite{Treangen2012} and the seeding strategy.

\begin{figure}[!htbp]
\hspace{1cm}
\includegraphics[width=0.4\textwidth, height=5cm]{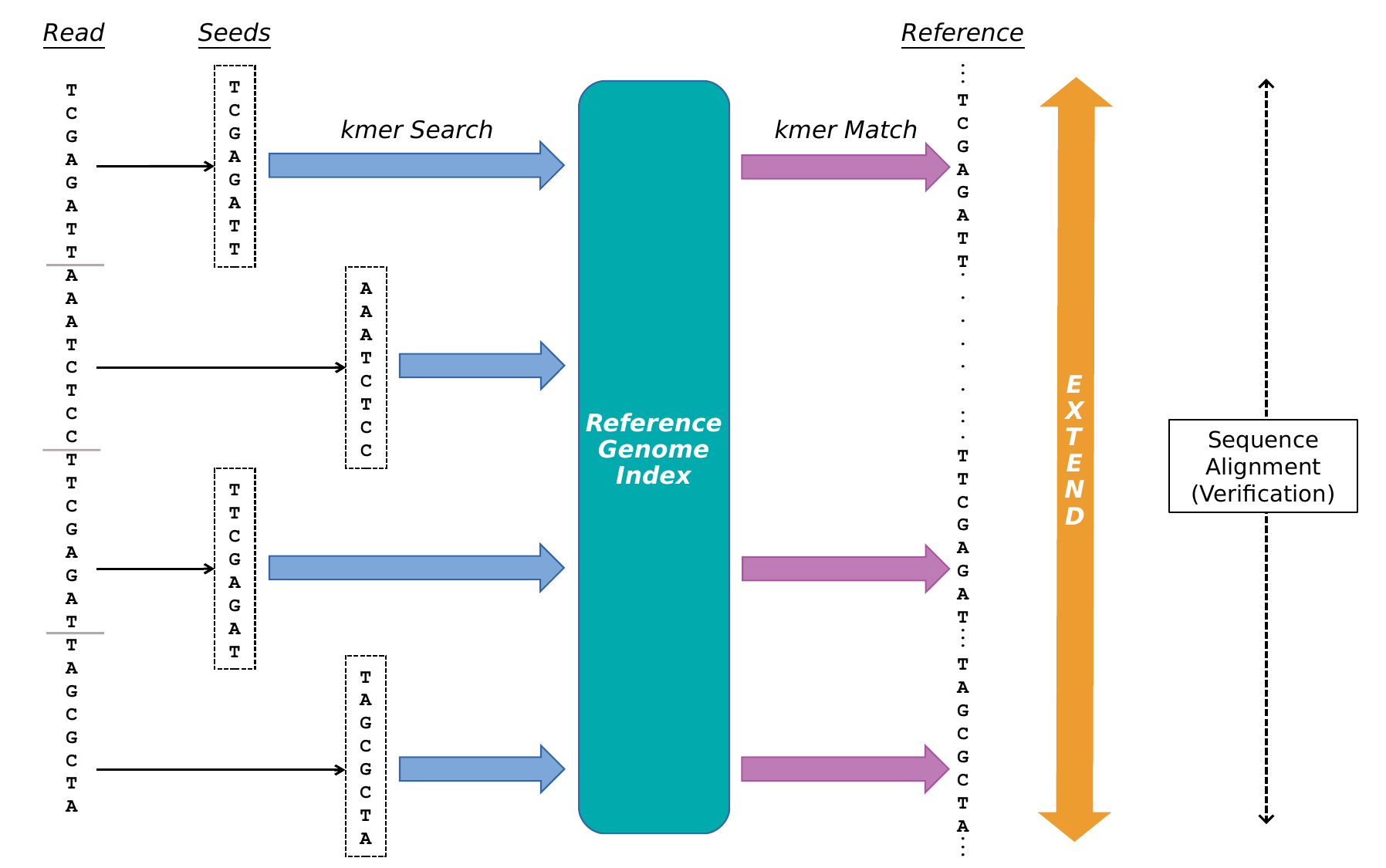}
\caption{High-level view of seed-and-extend paradigm.}
\label{fig:seed-n-extend}
\end{figure}

One of the most popular approaches to finding the possible locations that fit the read on the reference genome is the \textit{seed-and-extend} strategy~\cite{altschul1990basic}. This approach is rooted in the idea that two similar sequences share exactly or approximately matching substrings~(i.e., seeds, kmers)~\cite{ahmed2016comparison}. Consequently, once the matching locations of the shorter substrings, which are extracted from the read, are found on the reference genome, the seeds can then be \textit{extended} to form the candidate reference segment (Figure \ref{fig:seed-n-extend}). Since kmers are shorter than reads, the seeds may map to many locations, eventually creating multiple mapping locations for a single read.
\vspace{-2mm}

The presence of possible sequencing errors~\cite{shendure2008next} and base mutations restrict the use of an exact string matching algorithm, such as Hamming Distance~\cite{hamming1950error} or KMP~\cite{knuth1977fast}, for mapping the reads to the reference genome. Therefore, verifying the actual locations requires \textit{approximate string matching} techniques on strings with a predefined error threshold~\cite{alser2020accelerating}. Once all of the candidate locations are found by seeding, the read's most accurate location on the genome among all of its candidates is decided upon verification. Traditional practices generally lean towards using dynamic programming algorithms~\cite{alser2021technology} such as Needleman-Wunsch~\cite{needleman1970general}, Smith-Waterman~\cite{Smith1981}, and Levenstein Distance~\cite{levenshtein1966binary} to confirm that the distance between extended sequence segment from reference and the read is within the limits of an error threshold. However, dynamic programming solutions are computationally expensive with quadratic time and space requirements~(i.e., O($n^2$) for a sequence length of \textit{n}). Since seeding produces many candidate reference locations because of genomic repeats, performing alignment for every pair becomes a bottleneck. 

Due to the compute-intensive nature of dynamic programming solutions, the verification step creates a bottleneck for the entire read mapping procedure. Therefore, efficient amendments targeting this stage are expected to improve the entire process~\cite{alser2020accelerating}. The conspicuous approaches to address this problem would be either improving the algorithms in terms of time and space complexities~\cite{myers1999fast, zhang2000greedy} or reducing the workload on verification so that the pipeline visits verification as rarely as possible. In this work, our goal is to significantly reduce the execution time of read alignment by adopting the second approach and eliminating the candidate locations that exceed a predefined error threshold with fast and accurate pre-alignment filtering.
 
We propose GateKeeper-GPU, a fast and accurate general-purpose graphics processing unit~(GPGPU) pre-alignment filter for short read mapping to be performed before verification. GateKeeper-GPU is established on two key ideas:~1)~improving the filtering accuracy of GateKeeper~\cite{Alser2017}, which is a computationally lightweight pre-alignment filter, and~2)~employing massive parallelism provided by a large number of modern GPU threads for concurrent and quick examination of numerous sequence pairs. 
 
GateKeeper's original codebase is developed for field programmable gate array~(FPGA) devices. We opt for implementing it on general-purpose GPUs with the CUDA framework for several reasons. First, GPU has a large number of specialized cores~(i.e., 3584 CUDA cores in NVIDIA Geforce GTX 1080 Ti~\cite{cudacores}) that have smaller caches when compared to CPU cores; thus, its cores require lower frequency. This makes GPU a powerful candidate for processing the work for which the throughput is crucial~\cite{mittal2015survey}. Since pre-alignment filtering requires high throughput to complete as many comparisons as possible before verification, GPU is well-suited for this work. 
Second, although FPGA can provide more adjustable hardware designs, GPU is easier to install and use than FPGA because it does not require an in-depth understanding of the underlying hardware. Once installed, GPU can be used in a wide range of applications without extra configuration effort.
Third, GPU implementation is fully configurable at compile-time and runtime, such that the input parameters~(e.g., error threshold) can be changed without any alteration in the implementation.
On the other hand, the FPGA codebase requires changing the architecture design for each different parameter value. Therefore, we aim to provide easy usage and support for various platforms by implementing GateKeeper-GPU.

This paper makes the following \textbf{contributions}:
\begin{itemize}
\item We introduce GateKeeper-GPU, a new fast and accurate GPU-based pre-alignment filter with an improved GateKeeper algorithm and CUDA framework to facilitate a wider range of platform usage with GPGPUs.
\item We integrate GateKeeper-GPU with mrFAST to show its performance gain in a short read mapper with full workflow. This will hopefully serve as an example to embed GateKeeper-GPU to any short read mapping tool easily. 
\item We provide comprehensive analyses about the accuracy, filtering throughput, power, and resource utilization using two different device setups. We show that GateKeeper-GPU can accelerate the verification stage by up to $2.9\times$ and provide up to $1.4\times$ speedup for overall read mapping procedure when integrated with mrFAST \cite{alkan2009personalized}, in opposition to having no pre-alignment filter. Compared to the original GateKeeper~\cite{Alser2017}, GateKeeper-GPU produces up to $52\times$ less false accepts in candidate mappings.
\end{itemize}

\section{Background} 
\subsection{GateKeeper Filtering Algorithm}

GateKeeper~\cite{Alser2017} is the first FPGA-based pre-alignment filter that significantly improves the filtering speed. The algorithm is built upon Shifted Hamming Distance~(SHD)~\cite{xin2015shifted} and mainly consists of bitwise AND, XOR, and shift operations. Its simple and bitwise nature makes it applicable for hardware acceleration.

GateKeeper starts by encoding the read and its candidate reference segment in 2 bits (i.e., A = 00, C = 01, G = 10, T = 11) to prepare the bitwise representations of the strings. Then, it performs the XOR operation between bitvectors of the read and the reference segment to prepare the Hamming mask for exact match detection. Differentiating characters that depict mismatches are indicated as \textit{`1'} on the resulting bitvector, and matching characters are shown with \textit{`0'}. The approximate matching phase begins if the error threshold~(i.e., \textit{e}) is more than 0. In a loop of incrementing the error tolerance invariant \textit{k = $\{1,2,...,e\}$} one by one at each iteration, intermediate bitvectors are prepared by shifting read bitvector to the right and left by \textit{k} bits for deletion and insertion, respectively.
At the end of one iteration, the shifted bitvector undergoes XOR operation with reference segment bitvector to detect possible errors, thus every iteration produces two masks: one for deletion and one for insertion. Since the XOR operation signifies the difference in 2-bits per encoded character comparison, every two-bit is combined with bitwise OR to simplify the differences on individual bitvectors and reduce resource usage. 

The final stage of approximate matching is an AND operation on all \textit{2e+1} masks. Nevertheless, a \textit{0}-bit at a particular position is dominant, and AND will yield a \textit{0}-bit indicating a match even if all of the other bitvectors have a \textit{1}-bit value at that position, which signals a mismatch.
To compensate for this issue, the bitvectors are \textit{amended} before AND to turn short streaks of \textit{0}s into \textit{1}s considering these \textit {0}s are useless and do not represent an informative part in Hamming masks. For amendment, after the AND operation, the errors are counted by following a window approach with a look-up table. The comparison pair is rejected if the number of \textit{1s} as errors in the final bitvector exceeds the error threshold and is accepted if otherwise. For a step-by-step example of the GateKeeper algorithm's workflow, please refer to Sup. Figure S.1.  

\subsection{Unified Memory Architecture}
GPU provides a massive amount of parallelism to accelerate the compute-intensive work, which otherwise takes a long time with the CPU. The specialized cores of GPU process the data that physically reside on the device, thus, utilizing GPU as an accelerator inevitably requires moving the data from CPU to GPU memory via PCI-e bus. Because this creates an extra task for GPU when compared to CPU-only execution, adopting an effective memory management and data movement strategy plays a critical role. CUDA framework offers \textit{unified memory} for creating a virtual space, which GPU and CPU have access with a single pointer~\cite{harris_2018_unified}.

Unified memory does not remove the necessity of data movement altogether but maintains an automatic migration of data on-demand, providing data locality~\cite{harris_2018_unified}. Conceptually, the unified memory model fits GateKeeper-GPU in many aspects. For instance, the reference genome's designated segments are requested only when the mapper signals a potential match for the specific read in seeding, which creates an on-demand access situation. Likewise, it is sufficient to copy a single read only once to GPU memory for its multiple candidate reference segments. In our experience with different memory allocation methods at the development stage of GateKeeper-GPU, we also achieved more efficient overall execution with unified memory than with pinned memory format. Therefore, we adopted a unified memory model in GateKeeper-GPU.

Along with unified memory abstraction, the CUDA framework introduces \textit{memory advise} and \textit{asynchronous memory prefetching} features for optimized data migration during execution. Data access patterns of an object, such as preferred location, can be specified by memory advice so that the processor favors the optimal placement of the data for migration decisions. In addition, asynchronous data prefetching initiates the data transfer to the device before the data is used for minimizing the overhead caused by page faults in runtime~\cite{chien2019performance}. Prefetching is supported by Pascal and later architectures with CUDA 8 in GPU.

\subsection{Related Work}
Finding optimal solutions for the approximate string matching problem has been the focus of genome sequence studies due to the nature of data and its tremendous size. One of the tools built for this purpose, Edlib~\cite{vsovsic2017edlib}, is a CPU framework that calculates edit distance~(also known as Levenstein Distance~\cite{levenshtein1966binary}) by utilizing Myers' bitvector algorithm~\cite{myers1999fast} for optimal pairwise sequence alignment. Since Edlib finds the exact edit distance, we hold Edlib's global alignment results as the ground truth for our accuracy analysis.

The verification stage of read mapping tools determines the exact similarity between the given two sequences as read and candidate reference segment, therefore, the efforts on pre-alignment filtering prioritize quickly eliminating pairs with an apparent dissimilarity. Shifted Hamming Distance~(SHD)\cite{xin2015shifted} filters out highly erroneous sequence pairs with a bit-parallel and SIMD-parallel algorithm. MAGNET~\cite{alser2017magnet} addresses some of SHD's shortcomings, such as not considering leading and trailing zeros in bitmasks and consecutive bit counting for edit calculation, which are the main sources of high false accept rate. With a new filtering strategy, it improves the filtering accuracy by up to two orders of magnitude. Shouji~\cite{alser2019shouji}, on the other hand, takes a different approach to pre-alignment filtering. It starts by preparing a neighborhood map for identifying the common substrings between two sequences within an error threshold. Then, it finds non-overlapping common subsequences with a sliding window approach and decides on accepting or rejecting the pair according to the length of common subsequences. With an FPGA design, Shouji can reduce sequence alignment duration up to $18.8\times$ with two orders of magnitude higher accuracy than GateKeeper and SHD. SneakySnake~\cite{alser2020sneakysnake} solves approximate string matching by converting it to a single net routing problem~\cite{lee1976use}. This enables SneakySnake to provide acceleration with and without hardware support. SneakySnake also improves the accuracy of SHD, GateKeeper, and Shouji. Lastly, GenASM~\cite{cali2020genasm} is a hardware-accelerated approximate string matching framework for genome sequence analysis with a modified Bitap algorithm~\cite{baeza1992new, wu1992fast}. When utilized for pre-alignment filtering of short reads, it provides a $3.7\times$ speedup over Shouji while improving the accuracy.

\section{Methods}
The ultimate goal of GateKeeper-GPU is to accelerate the sequence alignment by quickly examining the sequence pairs with fast and accurate pre-alignment filtering, and deciding whether the computationally-expensive alignment is necessary for the genomic sequence pairs. There are four main steps. First, GateKeeper-GPU recognizes system specifications beforehand to allocate memory wisely and enable the extra features accordingly. Using compile-time variables, it calculates every other internal variable during execution via system configuration~(Section 3.1). Second, data buffers are allocated for an optimized data transfer between host and device~(Section 3.2). Third, reads and reference segments are preprocessed for the bitwise algorithm~(Section 3.3) and forth, filtration is applied on each sequence pair by GateKeeper-GPU kernel with an improved GateKeeper algorithm~(Section 3.4). Lastly, we integrate GateKeeper-GPU into mrFAST~\cite{alkan2009personalized} to observe the practical performance gains~(Section 3.5). 

\subsection{System Configuration}
GateKeeper-GPU is designed to require the least amount of user interaction and effort. Because of the disadvantages of dynamic kernel memory allocation of CUDA, read length and error threshold should be specified at compile time. It should be emphasized that the read length variability across different Illumina data sets is low and most of the data sets include 100bp or 150bp reads, therefore the tool recompilation for each dataset is not a frequent necessity.
GateKeeper-GPU starts with configuring the system's properties, such as device compute compatibility since some of the features, like data prefetching, are limited to the compute capability of the system.

Applying the GateKeeper algorithm for a read-reference segment pair is called \textit{filtration}. Since CUDA provides many massively parallel threads, each filtration is performed by a single CUDA thread to have the least possible dependency between the threads for high filtering throughput. Each thread uses the reserved stack frame for the temporary data, such as bitmasks. Before the filtering step, GateKeeper-GPU calculates the approximate memory load of filtration on a thread~(i.e., thread load) using the compilation variables of read length and error threshold. It retrieves the free global memory size of the GPU along with some other GPU parameters, calculates the number of CUDA thread blocks, and the possible number of filtrations for one kernel call~(i.e., batch size) to fully utilize GPU for boosting performance. Since data transfers between the device and host are expensive, the configuration step ensures that the batch size is maximized to keep the total number of transfers minimal. In the multi-GPU model, the batch size is equal for all devices to ensure a fair workload. In this way, we adjust the kernel parameters of efficient GPU usage for the device model and memory status without the user’s concern.

\subsection{Resource Allocation}
The main algorithm consists of simple bitwise operations. Distributing the workload of one filtration between multiple threads would introduce overheads emerging from inter-thread dependencies rather than speeding up the process. Therefore, each thread runs kernel function for a single filtration with the least dependency possible. 

GateKeeper produces \textit{2e + 1} Hamming masks to store the intermediate bitvectors for indels~(i.e., insertions and deletions) where \textit{e} is the error threshold. Since CUDA dynamic memory allocation inside device functions is restricted and slow, we use fixed-size unsigned integer arrays for bitmasks in the kernel. For this reason, GateKeeper-GPU requires read length and error threshold at compile time. Each thread uses the reserved stack frame in thread-local memory for bitmasks, which is cached in unified L1/Texture and L2 cache for Pascal architecture; in L1 and L2 cache for Kepler architecture~\cite{programmingGuide}. All of the constant variables and the look-up table~(LUT) for the amending procedure are stored in the constant memory.

We utilize unified memory for the read buffer and the reference for providing simplicity and on-demand data locality. As the number of filtration operations per batch is determined in the system configuration stage, a read buffer is created in unified memory together with candidate reference indices corresponding to the reads. Since the number of candidate reference locations is unknown until seeding, batch size does not limit candidate reference location count, thus there is no predefined value of reference segment count per read.     

\subsection{Preprocessing}
Once the system properties are recognized and parameters are adjusted accordingly, GateKeeper-GPU starts to prepare the reads and the reference for filtration. The main portion of preprocessing involves encoding the strings to 2-bit representations for bitwise operations. Since each character takes up two bits after encoding, a 16-character window is encoded into an unsigned integer (i.e., one word), thus a 100bp read is represented as seven words in the system.    

GateKeeper is designed for DNA strings, hence it only recognizes the characters \textit{`A', `C', `G'} and \textit{`T'}. Occasionally, the reads or the reference may contain the character \textit{`N'} that signals for an unknown base call at that specific location of the sequence. Because GateKeeper does not recognize the character \textit{`N'}, GateKeeper-GPU directly passes those sequence pairs from the filter without applying filtration steps as a design choice for two reasons: first, supporting an extra character requires expanding the encoding to 3-bit representations which unnecessarily increases the complexity and memory footprint for this rare occasion; second, since these unknown bases add extra uncertainty, leaving the decision to verification increases credibility. 

Assigning the encoding job to either the host~(i.e., CPU) or the device~(i.e., GPU) creates advantages and disadvantages from different perspectives. Encoding in the host and copying the encoded strings to the device is cost-effective in data transfer since encoding compresses the strings into smaller units. However, processing in the host can be stagnant even if it is multi-threaded compared to massively parallel processing in the device. Conversely, allowing each device to encode the strings for their own operations adds extra parallelism to the whole execution while reducing the efficiency in transfer time. We provide GateKeeper-GPU in both versions and analyze the effect of the processor in encoding on overall performance, which will be discussed further in the Evaluation section.

In most of the read mappers' workflow, first, the possible reference locations are found, the read is verified, and then the same operations are carried out for the next read. Multi-threaded mappers allow the processing of several reads at a time. Nevertheless, it is necessary to utilize the maximum number of threads possible per kernel call for an effective GateKeeper-GPU execution by fully employing the GPU resources. Therefore, many read and their candidate reference segments are prepared and batched in the host for a single kernel before verification, depending on the pre-calculated batch size.

\subsection{GateKeeper-GPU Kernel}
When the buffers are ready for execution, we set the usage patterns of the buffers in terms of caller frequency about the host or the device by CUDA memory advice API. Since the kernel utilizes the buffers more than CPU functions until the next batch of reads is processed, the preferred location of the data is set to be the GPU device for the input buffers for the kernel. According to the assigned memory advice for each buffer, the data arrays are prefetched asynchronously ahead of the kernel function to provide an early start for data migration from host to device, and mapping data to the device's page tables before kernel accesses~\cite{programmingGuide}. Since there is more than one buffer for prefetching, each buffer is asynchronously submitted to a different stream. Memory advice mechanism and prefetching are supported for compute capability $6.x$ and later, therefore these actions are skipped for lower CUDA compute capabilities.

After prefetching, the kernel function is executed with the parameters for the number of blocks and threads that are previously set in the configuration stage. The kernel performs the complete set of operations for a single filtration, starting with encoding the sequences if they are not encoded in the preprocessing stage. 

Due to architectural differences between FPGA and GPU, some alterations are applied to maintain the GateKeeper algorithm in C-derived device functions in GPU. In contrast to FPGA's specialty in bitwise operations, GPU does not support bitvectors in arbitrary sizes. An encoded 100bp read can be represented as a 200-bit long register in FPGA, whereas GPU allocations are limited with the word size of the system, thus the encoded read becomes an array of 7 words. Additionally, logical shift operations produce incorrect bits between array's elements. For correcting these bits, we apply \textit{carry-bit} transfers. The correction procedure must be performed for each bitwise shift, such that there are \textit{2e} shifts and \textit{2e} carry-bit operations~(insertion and deletion masks each require \textit{e} operations) in the filtration with an error threshold of \textit{e}. 

\begin{figure}[!htbp]
\centering
\includegraphics[width=0.35\textwidth]{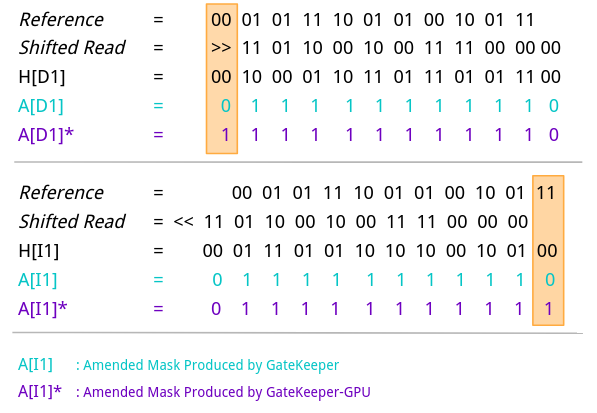}
\caption{Strategy for improving leading and trailing 0 bits. \textit{Reference} and \textit{Shifted Read} show bitvectors of candidate reference segment and shifted read, respectively; \textit{H} and \textit{A} represent Hamming mask and amended mask.}
\label{fig:imp_mini}
\begin{minipage}{0.49\textwidth}
\vspace*{0.1cm}
\end{minipage}
\end{figure}

Bitwise shift operations leave the most significant or the least significant bits vacant, depending on and opposite to the shift direction. Even though these bits should be $1s$ on the final bitvector signaling for errors, the final AND operation on all masks hide these errors since the corresponding bits are $0$ in the shifted masks. MAGNET~\cite{alser2017magnet} previously addressed this issue by solving it with a combination of steps. In order to uncover these possible errors on leading and trailing parts, we add a single OR operation to turn the excess bits into $1s$ after preparing amended masks. In this way, we ensure that the leading and trailing bits are $1$ even if the XOR operation for the Hamming mask converts them into $0$. Figure \ref{fig:imp_mini} shows how the new amended mask covers the leading and trailing $0$, which is missed by GateKeeper~\cite{Alser2017}. 
\vspace{-2mm}

\begin{figure}[!htbp]
\centering
\includegraphics[width=0.5\textwidth]{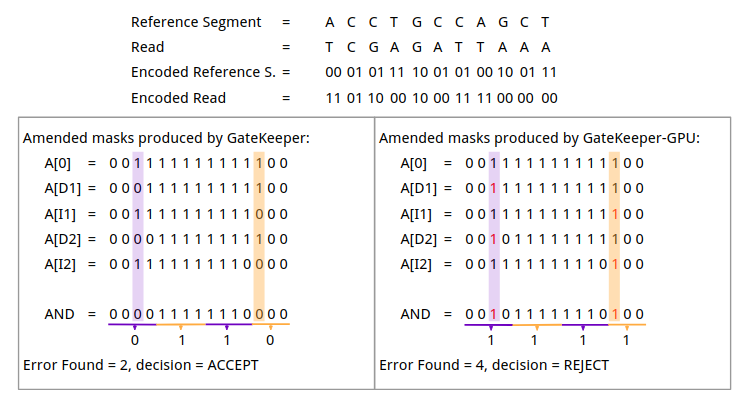}
\caption{Amended masks produced by GateKeeper \cite{Alser2017} and GateKeeper-GPU.}
\label{fig:imp_big}
\end{figure}

As a result of this modification, GateKeeper-GPU can reject some of the read and reference segment pairs that exceed the error threshold, which GateKeeper falsely accepts. Figure \ref{fig:imp_big} shows the amended masks produced by GateKeeper and GateKeeper-GPU for the same sequence pair for error threshold \textit{e} = 2. This improvement enables GateKeeper-GPU to make the correct decision for rejecting the pair, whereas GateKeeper falsely accepts the pair.

It should be emphasized that GateKeeper-GPU does not perform seeding, and it is not a kmer filter. It relies on the candidate reference locations reported by the mapper and performs filtration only on the read and reference segment pairs, which are expected to match with the mapper's high confidence. Further, GateKeeper-GPU does not calculate but \textit{approximates} the edit distance between pairs for fast filtration. The verification performs the exact edit distance calculation, and GateKeeper-GPU acts as an intermediate step in preparation for verification.

\subsection{Adaptation to mrFAST Workflow}
We integrate GateKeeper-GPU into mrFAST to evaluate its performance on the whole genome scale, as briefly illustrated in Sup. Figure S.2. GateKeeper-GPU can be adapted to any short read mapping tool that uses seed extension, and in this section, we present the necessary adjustments. We begin with encoding and loading the reference into the unified memory using multithreading with OpenMP. While encoding, the locations of \textit{`N'} bases on the reference genome are also recorded since the segments containing this character will not be evaluated in the filtering stage. 

In the original workflow of mrFAST, candidate locations of a single read are found and verified, mapping information of the read is recorded, then the next read is processed. Since GateKeeper-GPU requires batching, we fill the buffers with multiple reads and their candidate location indices with partial multicore support. The number of candidate locations of a particular read cannot be anticipated before seeding, therefore there are two factors that dynamically control the size of buffers to ensure that GPU utilization is optimized without exhausting the resources: the number of filtration pairs that is calculated by the system configuration unit~(Section 3.1) and the number of reads allowed per batch. The number of reads is predetermined and in our experience with different values~(Table \ref{tab:maxnumreads}), we find that $100,000$ reads yield the best results for mrFAST. We observe that using $100,000$ reads per batch decreases the overall runtime, durations of the kernel, and filter since the number of transfers between the host and the device is minimized. Still, the maximum number of reads is a parameter for execution and can be modified easily according to the desired mapper or execution.

\begin{table}[!htbp]
\caption{Effect of the maximum number of reads processed per batch on time~(seconds).}
\label{tab:maxnumreads}
\begin{adjustbox}{max width=0.49\textwidth}
\Large
\begin{tabular}{r|rrrr|rrrr}
\multicolumn{1}{c}{}                          & \multicolumn{4}{c}{\textbf{Encoding in Host}}                                                                                                            & \multicolumn{4}{c}{\textbf{Encoding in Device} }                                                                                                           \\
\multicolumn{1}{c|}{ \textit{Max. \# Reads} } & \multicolumn{1}{c}{\textit{Overall} } & \multicolumn{1}{c}{\textit{Encode }} & \multicolumn{1}{c}{\textit{Kernel }} & \multicolumn{1}{c|}{\textit{Filter} } & \multicolumn{1}{c}{\textit{Overall} } & \multicolumn{1}{c}{\textit{Copy} } & \multicolumn{1}{c}{\textit{Kernel }} & \multicolumn{1}{c}{\textit{Filter} }  \\ 
\hline \rule{0pt}{2ex}
100                                           & 3,041.52                              & 109.54                               & 102.55                               & 212.17                                & 2,944.59                              & 100.19                             & 105.39                               & 187.58                                \\
1000                                          & 1,446.58                              & 105.99                               & 92.72                                & 114.61                                & 1,335.20                              & 77.55                              & 97.20                                & 116.29                                \\
10,000                                        & 1,325.95                              & 109.14                               & 80.37                                & 92.99                                 & 1,322.96                              & 84.45                              & 83.22                                & 92.29                                 \\
100,000                                       & 1,275.66                              & 103.13                               & 77.45                                & 88.96                                 & 1,215.25                              & 75.19                              & 82.37                                & 91.17                                \\
\hline 
\end{tabular}
\end{adjustbox}
\begin{minipage}{0.49\textwidth}
\vspace*{0.2cm}
{\scriptsize
These measurements were recorded during the mapping of Chromosome 1. Kernel: Total filtering time measured by CUDA API, Filter: Total filtering time measured from the host side.}
\end{minipage}
\end{table}


Once the data transfer buffers are filled, the kernel function is called with the previously calculated number of blocks and threads parameters. Each thread executes a single comparison, starting with extracting the relevant reference segment based on the index. The result of the filtering decision as `1'~(\textit{accept}) or `0'~(\textit{reject}), and the approximated edit distance are written back on the result buffers in the unified memory. 

The only synchronization point for threads is after the completion of the filtering step for one batch. Since the host and device use a common pointer for buffers in the unified memory model, the threads' job needs to be completed in order for verification to obtain filtering results. The pairs that pass the filter are verified, and mrFAST continues with further steps to report the mapping information.

\section{Experimental Methodology}
\subsection{Data sets} 
We run whole genome tests on mrFAST with one real and two simulated data sets. We obtained the biological data sets from the 1000 Genomes Project Phase I~\cite{10002012integrated} and produced simulated reads using Mason~\cite{holtgrewe2010mason} at different read lengths. In all of our tests, we use the reference GRCh37 produced by the 1000 Genomes Project~\cite{10002015global}. In order to have a fair comparison for accuracy, we use the same data sets from GateKeeper~\cite{Alser2017} for the accuracy tests. 

For filtering throughput and accuracy analyses, the datasets contain 30 million read and candidate reference segment pairs, seeded by mrFAST, using biological short read sets~(downloaded from European Nucleotide Archive) with different error threshold values. Additionally, we generate read and candidate reference segment pairs using two of the state-of-the-art mapping tools, Minimap2~\cite{li2018minimap2}, and BWA-MEM~\cite{li2013aligning} for accuracy analyses. For Minimap2, we extract the pairs just before the first chaining function ($mm\_chain\_dp$) since this is the first dynamic programming part and gather the samples in 11 different sets for error threshold from 0 to 10, each containing 30 million pairs. For BWA-MEM, we extract the pairs before the final global alignment call (\textit{ksw\_global2}). BWA-MEM generates much fewer pairs than 30 million at this point, therefore the data set size for each error threshold~(\textit{e} = {0, \ldots, 10}) is different.  
By using Edlib's global alignment mode, we prepare the filtering status of data sets to maintain consistency on ground truth. In all of the experiments, the maximum error threshold is $10\%$ of the read length. Please refer to Sup. Table S.1 for the details about the data sets.
\subsection{Experimental Setup}
We run our performance experiments in two different setups. \textit{Setup\_1} includes a 2.30GHz Intel Xeon Gold 6140 CPU with 754G RAM. There are 8 NVIDIA GeForce GTX 1080 Ti GPUs~(Pascal architecture), each with 10GB global memory connected to this processor. CUDA compute capability of the devices is 6.1 with CUDA driver v10.1 installed. Data transfer between the host and devices is managed by PCIe generation 3 with 16 lanes. In \textit{Setup\_2}, we use a 3.30GHz Intel Xeon CPU E5-2643 0 processor with 256G RAM. 4 NVIDIA Tesla K20X GPGPUs are connected to this host. Each device has 5GB global memory and the data transfer is maintained via PCIe generation 2 with 16 lanes. CUDA compute capability of the devices is 3.5 with CUDA driver v10.2. Tesla K20X has Kepler architecture, therefore data prefetching is not supported in Setup\_2. In all of our tests, we enable persistence mode in GPU to keep the devices initialized.

\subsection{Filtering Throughput Evaluation}
We run our throughput analysis on the datasets containing read and reference segment pairs. Each dataset collected for this purpose includes 30 million pairs in total. We calculate the runtime taken for the filtration of a single pair out of 30 million, then determine the total number of pairs that can be filtered in 40 minutes to have a fair throughput comparison with the other tools. 

We report two different time measurements for throughput analysis: \textit{kernel time} and \textit{filter time}. \textit{Kernel time} denotes the time taken only by GPU devices and we record this time using CUDA Event API. Since GateKeeper-GPU uses batched kernel calls, we add all kernel times in execution and report the sum. \textit{Filter time} represents the total time spent for filtering, including host operations such as data transfer and encoding the sequences. Therefore, we measure filter time from the host's perspective. For both of these measurements, we also provide different results for encoding the pairs by the host~(CPU) and the device~(GPU). In multi-GPU throughput analysis, kernel time represents the time of the device, which takes the longest time to complete among all other active devices. In all of our timing experiments, we run the tests ten times and report the arithmetic mean to minimize the effect of random experimental errors on results. 

We compare GateKeeper-GPU's filtering throughput with its CPU version comprehensively and make a brief comparison with its FPGA version~\cite{Alser2017}. To maintain fairness as much as possible, we implement GateKeeper-CPU in a multicore fashion and report the results of 12 cores.   

\subsection{Accuracy Evaluation} 
In our accuracy analyses, we consider Edlib's~\cite{vsovsic2017edlib} edit distance as the ground truth and calculate the edit distance between the read and reference segments by using Edlib's global alignment. According to the edit distance, we produce a filtering status for each pair as \textit{reject} if the edit distance is larger than the threshold or \textit{accept} if otherwise. Throughout accuracy experiments, we analyze \textit{false accept}, \textit{false reject}, and \textit{true reject} counts. A false accept represents a read and reference segment pair that Edlib rejects because of exceeding the error threshold, but is accepted by the filter. On the contrary, a false reject case is a valid pair with fewer errors than the threshold but is rejected by the filter. True rejects are the pairs that are rejected by both Edlib and GateKeeper-GPU.

We have two approaches for testing the accuracy of GateKeeper-GPU. First, we record the filtering status for the datasets described in Section 4.1 and compare the results with Edlib. Since GateKeeper-GPU gives a direct pass to the pairs that contain unknown base call characters~(\textit{`N'}), to be able to observe the actual accept and reject counts, we exclude these pairs from the tests and report the comparison with Edlib accordingly. For the sake of simplicity, we will call these pairs \textit{undefined} for the rest of the work. 

Second, we compare GateKeeper-GPU with the original GateKeeper FPGA implementation, SHD, MAGNET, Shouji, and SneakySnake in terms of false accept and false reject counts. We denote the original GateKeeper implementation as `GateKeeper-FPGA' for labeling. We choose the highest-edit and lowest-edit profile datasets of three different read lengths for these tests. For 100bp, the lowest-edit containing data set is prepared by mrFAST's seeding error threshold \textit{e =}~2, and the highest-edit data set is curated by error threshold \textit{e =}~40. Likewise, mrFAST error thresholds for the lowest-edit and the highest-edit profile data sets are \textit{e =}~4 and \textit{e =}~70 for 150bp; \textit{e =}~8 and \textit{e =}~100 for 250bp, respectively. Because the other tools do not have distinguishing mechanisms for undefined pairs, to maintain a fair comparison in this test series, we include these pairs in GateKeeper-GPU's results and report the numbers accordingly. 

\subsection{Whole Genome Evaluation} 
We evaluate GateKeeper-GPU's performance and accuracy in the full workflow of a read mapping tool by integrating GateKeeper-GPU into mrFAST. GateKeeper was also integrated into mrFAST for testing. Therefore, testing GateKeeper-GPU as a part of mrFAST enables us to briefly compare it with the original work. 

We use the same datasets that GateKeeper~\cite{Alser2017} uses in whole genome comparison experiments: one real 100bp data set and one simulated 300bp data set~(i.e., sim\_set\_1). We add another simulated 150bp data set~(i.e., sim\_set\_2) analysis to GateKeeper-GPU's results. We collect the following metrics from alignment for evaluation: the number of mappings, the number of mapped reads, the total number of candidate mappings, the total number of candidate mappings that enter verification, time spent for verification, time spent for preprocessing before pre-alignment filtering, and total kernel time spent for running GateKeeper-GPU. We use a single GPU in these experiments and provide results for encoding in both device and host design. 
\subsection{Resource Utilization and Power Evaluation} 
Warp occupancy is one of the important indicators when evaluating the kernel performance of a CUDA application. A \textit{warp} is a group of 32 adjacent threads in a block, and Streaming Multiprocessor~(i.e., SM) schedules the same instruction for the threads in a warp~\cite{achieved_occupancy}. Warp occupancy is the ratio of active warps over the maximum supported number of warps for SM. The number of warps, blocks, registers per SM, and shared memory determines theoretical occupancy. With these conditions, the maximum number of registers per thread is 32 for $100\%$ occupancy while using all threads in a warp. GateKeeper-GPU does not utilize shared memory, and we opt to maximize the number of warps and blocks to maintain a high throughput. Using the CUDA occupancy calculator~\cite{cudaoccupancycalculator}, we calculate GateKeeper-GPU's theoretical warp occupancy and present a comparison with the achieved occupancy value. 

We perform profiling experiments on the 100bp and 250bp data sets with error thresholds \textit{e = }4 and \textit{e = }10, respectively, by using CUDA command-line profiler nvprof \cite{nvprof}. By conducting system profiling and metrics analyses, we provide GateKeeper-GPU's power consumption, warp execution efficiency, and multiprocessor activity.

\section{Evaluation}
\subsection{Accuracy Analysis}

\subsubsection{Accuracy of GateKeeper-GPU with respect to Edlib}
In the first phase of accuracy analyses, we evaluate the accuracy of GateKeeper-GPU against Edlib using data sets with three different read lengths. We exclude the undefined pairs for both tools in this test series to indicate the actual numbers of accepts and rejects without skipping filtration. We perform the experiments on the data sets that include the pairs with $5\%$ of their length error allowance by mrFAST. With this rule being set; Set\_3, Set\_6, and Set\_10 contain reads and mrFAST's candidates for error thresholds respectively 5, 6, and 12 (Sup. Table S.1).
We perform experiments on these data sets with filtering error thresholds from $0\%$ to $10\%$ of their corresponding read length. Additionally, we carry out accuracy tests with potential mappings of Minimap2, as described in detail in Section 4.1.        
\begin{figure}[!htbp]
  \centering
  \begin{minipage}[b]{0.45\textwidth}
    \includegraphics[width=\textwidth]{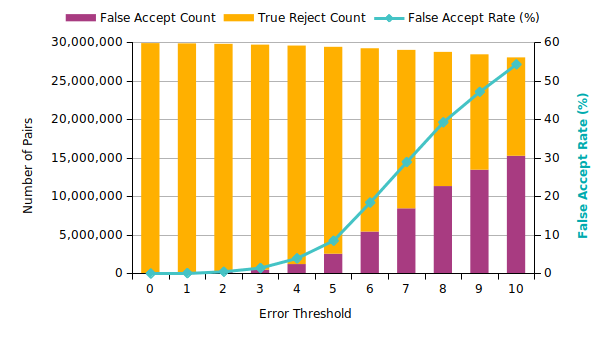}
    \caption{False accept analysis - 100bp.}
    \label{fig:acc_e5_100bp_stacked}
  \end{minipage}
\end{figure}


Compared with Edlib's global alignment results, GateKeeper-GPU's false reject count is always $0$ for all data sets, indicating that it never rejects a true extendable read and reference segment pair. For mrFAST's potential mappings, Figure \ref{fig:acc_e5_100bp_stacked} illustrates the number of false accepts and true rejects and the false accept rate as a percentage of the number of false accepts over the number of rejected pairs by Edlib, with respect to the error threshold. The detailed results of these experiments and the figures for the remaining data sets are available in Sup. Material IV.

Considering the results of these experiments, we make the following observations regarding the accuracy of GateKeeper-GPU:~1)~Up to $\sim$3\% error thresholds of all read lengths, GateKeeper-GPU can correctly reject more than $90\%$ of the mappings with less than $10\%$ of false accept ratio and with no false reject.
2)~Even though the efficiency of filtering decreases when the error threshold increases, filtering still continues to serve without a steep drop in efficiency for the largest error threshold allowed~($10\%$). 3)~With increased read length, the false accept rate increment and filtering efficiency decrease become sharper. Furthermore, GateKeeper-GPU can filter out all dissimilar sequence pairs produced by Minimap2 with no false accepts for error threshold \textit{e} = 0~(Sup. Table S.5) and can present a true reject rate of up to 98\% on BWA-MEM's candidate mappings~(Sup. Table S.6).

\subsubsection{Comparison with Other Pre-alignment Filters}
We compare GateKeeper-GPU's results with five other filtering tools: its original FPGA GateKeeper implementation, SHD, MAGNET, Shouji, and SneakySnake with respect to Edlib's ground truth in the second phase of accuracy analyses. For this purpose, we use low-edit profile~(Set\_1, Sec\_5, and Set\_9) and high-edit profile datasets~(Set\_4, Sec\_8, and Set\_10), previously described in Section 4.1 and presented in Sup. Table S.1 in detail. We perform experiments on these data sets with the filtering error thresholds from $0\%$ to $10\%$ of their corresponding read length.  We retrieved other tools' false accept and reject counts from the Sup. Material of the Shouji manuscript. To maintain a fair comparison in this test series, we include the undefined pairs and mark these pairs as falsely accepted in GateKeeper-GPU's results where necessary.

\begin{figure}[!htbp]
  \centering
  \begin{minipage}[b]{0.45\textwidth}
    \includegraphics[width=\textwidth]{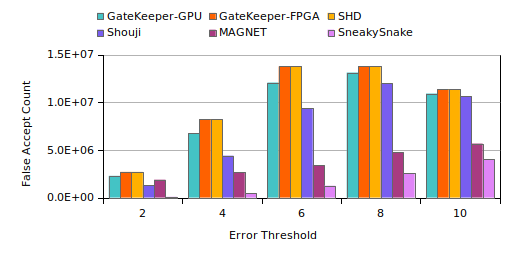}
    \caption{False accept comparison for Set\_1 with a read length of 100bp and the number of undefined pairs is 28,009.}
    \label{fig:FP_e2_100bp}
  \end{minipage}
\end{figure}

Figure \ref{fig:FP_e2_100bp} and Sup. Figures S.7 to S.11 demonstrate the number of falsely accepted pairs by different tools across the same datasets with varying error thresholds. Even though GateKeeper-GPU has undefined pairs in its false accept count, we see that it has a less false accept rate in most of the cases and can produce up to $52\times$ less number of false accepts~(Table S.10, \textit{e} = 9) when compared to original GateKeeper-FPGA and SHD. We observe that both GateKeeper-FPGA and SHD completely stop filtering in high error thresholds of high-edit profile datasets and accept all pairs~(30 million). In contrast, GateKeeper-GPU still functions in those cases and continues to decrease the number of potential mappings correctly. This suggests that the small modifications made on the GateKeeper algorithm for leading and trailing parts of the bitvectors, explained in detail in Section 3.4, improve the accuracy and help GateKeeper-GPU keep its consistency without completely letting loose filtering in high error thresholds.

In all the tests, SneakySnake and MAGNET have the lowest number of false accepts. However, we notice that MAGNET produces some false accepts for exact matching when the error threshold is 0 as seen in Sup. Figure S.8, and generates false rejects in some cases where GateKeeper-GPU and other filters do not have false rejects. GateKeeper-GPU rarely produces false accepts in the exact matching, and the number of false accepts in these cases is always lower than MAGNET. Finally, for datasets with read lengths 150bp and 250bp, GateKeeper-GPU and Shouji have similar false accept rates, but Shouji's false accept count is less than GateKeeper-GPU. More detailed results of these tests are available in Sup. Material V.

\subsection{Filtering Throughput Analysis}
To assess GateKeeper-GPU's filtering throughput, we perform experiments on datasets Set\_3, Set\_7, and Set\_11. With respect to kernel time~(i.e., kt) and filter time~(i.e., ft), we calculate the filtering throughput of GateKeeper-GPU. We report the results with filtering error thresholds \{2, 5\}, \{4, 10\}, and  \{6, 10\} for the datasets with read lengths 100bp, 150bp, and 250bp respectively. Table \ref{tab:100bp_filtering_throughput} contains the filtering throughput of GateKeeper-CPU and GateKeeper-GPU with different values for encoding in device and host designs in terms of billions of filtrations in 40 minutes. For GateKeeper-CPU, kernel time represents the time exclusively spent by the function that contains the GateKeeper algorithm. We provide GateKeeper-CPU's results for single-core and 12-core experiments, but we make comparisons based on 12-core results. Please refer to Sup. Material VI for the detailed results of these experiments.

Considering filter time results in Setup\_1, GateKeeper-GPU can filter up to $3\times$ and $20\times$ more pairs than 12-core GateKeeper-CPU, with single and 8 GPUs, respectively (Sup. Table S.15, device\_encoded, 250bp, \textit{e} = 10). In terms of only kernel time, GateKeeper-GPU's filtering throughput can reach up to $72\times$ and $456\times$ more than 12-core GateKeeper-CPU, with single and 8 GPUs, respectively~(Sup. Table S.15, host\_encoded, 250bp, \textit{e} = 6). In Setup\_2 with a single GPU, GateKeeper-GPU can filter up to $2\times$ and $16\times$ more pairs than 12-core GateKeeper-CPU with respect to filter time and kernel time (Sup. Table S.15, device\_encoded, 250bp, \textit{e} = 10).

\begin{table}[!htbp]
\caption{Filtering throughput for 100bp sequences.}
\label{tab:100bp_filtering_throughput}
\begin{adjustbox}{max width=0.49\textwidth}
\Large
\begin{tabular}{lcc|rr|rr|rr}
                                                        &                                         & \multicolumn{1}{c}{\textbf{} }  & \multicolumn{2}{c}{\textbf{GateKeeper-CPU} }                                    & \multicolumn{2}{c}{\textbf{Device-encoded} }                               & \multicolumn{2}{c}{\textbf{Host-encoded} }                                 \\  
\cline{4-9}   
\multicolumn{1}{c}{}    \rule{0pt}{3ex}                                 &                                         & \multicolumn{1}{c}{\textit{e}~} & \multicolumn{1}{r}{\textit{1-Core } } & \multicolumn{1}{r}{\textit{12-Cores } } & \multicolumn{1}{r}{\textit{1-GPU } } & \multicolumn{1}{r}{\textit{8-GPU} } & \multicolumn{1}{r}{\textit{1-GPU } } & \multicolumn{1}{r}{\textit{8-GPU}}  \\ 
\cline{3-9}
\multirow{4}{*}{Setup\_1}                     &  \rule{0pt}{2ex} \multirow{2}{*}{kt}                     & 2                               & 0.7                                   & 7.2                                     & 244.8                                & 1,189.8                             & 476.8                                & \textbf{3,198.4}                    \\
                                                        &                                         & 5                               & 0.4                                   & 3.9                                     & 150.8                                & 1,041.4                             & 249.3                                & \textbf{1,684.7}                    \\ 
\cline{3-9}
                                                        &  \rule{0pt}{2ex} \multirow{2}{*}{ft}                     & 2                               & 0.6                                   & 6.5                                     & 7.7                                  & \textbf{39.2}                       & 3.0                                  & 14.4                                \\
                                                        &                                         & 5                               & 0.4                                   & 3.7                                     & 7.6                                  & \textbf{37.8}                       & 2.9                                  & 14.2                                \\ 
\cline{2-9}
\multicolumn{1}{r}{\multirow{4}{*}{Setup\_2}}  \rule{0pt}{2ex}& \multicolumn{1}{r}{\multirow{2}{*}{kt}} & 2                               & 0.7                                   & 5.5                                     & 41.1                                 & NA                                  & \textbf{72.2}                        & NA                                  \\
\multicolumn{1}{r}{}                                    & \multicolumn{1}{r}{}                    & 5                               & 0.3                                   & 3.0                                     & 29.1                                 & NA                                  & \textbf{42.0}                        & NA                                  \\ 
\cline{3-9}
\multicolumn{1}{r}{}             \rule{0pt}{2ex}                        & \multicolumn{1}{r}{\multirow{2}{*}{ft}} & 2                               & 0.6                                   & 4.9                                     & \textbf{6.1}                         & NA                                  & 2.7                                  & NA                                  \\
\multicolumn{1}{r}{}                                    & \multicolumn{1}{r}{}                    & 5                               & 0.3                                   & 2.8                                     & \textbf{5.7}                         & NA                                  & 2.7                                  & NA                                  \\
\cline{3-9}
\end{tabular}
\end{adjustbox}
\begin{minipage}{0.49\textwidth}
\vspace*{0.2cm}
\scriptsize
Filtering throughput is calculated wrt. kernel time (kt) and filter time (ft), in terms of billions of pairs in 40 minutes. The highest filtering throughput within the row is in bold font. \textit{e = }error threshold.
\end{minipage}
\end{table}

Although in some cases, 12-core GateKeeper-CPU exhibits similar performance to single device GateKeeper-GPU with encoding in host option in terms of filter time, the biggest advantage of GPU implementation over CPU implementation is having a constant performance when the error threshold increases. While the filter time remains constant in both device-encoded and host-encoded GateKeeper-GPU, the growth in GateKeeper-CPU’s filter time is almost linear with increasing error threshold. Hence, GateKeeper-GPU performs better with high error thresholds.

\begin{figure}[!htbp]
  \centering
  \begin{minipage}[b]{0.45\textwidth}
    \includegraphics[width=\textwidth]{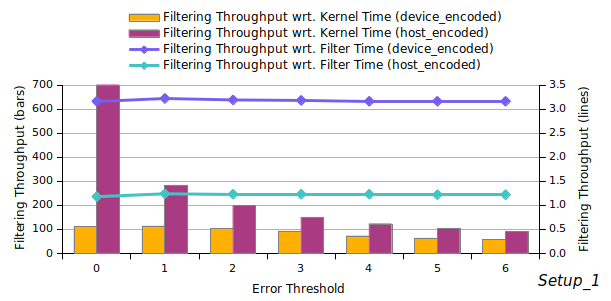}
  \end{minipage}
  \begin{minipage}[b]{0.46
  \textwidth}
    \includegraphics[width=\textwidth]{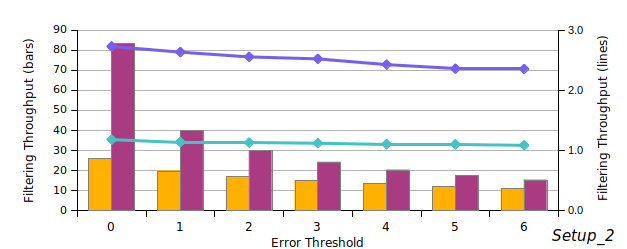}
  \end{minipage}
  \vspace{0.2cm}
  \caption{\scriptsize Effect of the encoding actor~(device or host) on filtering throughput (millions of filtrations per second) of single-GPU GateKeeper-GPU for 100bp reads. Filtering throughput is calculated with respect to kernel time: bars, and filter time: lines.}
  \label{fig:thr_encodingfact}
\end{figure}

GateKeeper-GPU's performance changes depending on the encoding actor~(as host or device) as Figure \ref{fig:thr_encodingfact}, and Sup. Figures S.13 and S.14 show. We notice that encoding in the host leads to higher throughput in all three different sequence lengths when we consider kernel time (depicted by bars in the figures). Especially in lower error thresholds, the difference between host-encoded and device-encoded throughput values is significant. On the other hand, it turns into the opposite situation when filter time (depicted by lines in figures) is considered. From these observations, we conclude that the encoding procedure creates a bottleneck for the workflow of GateKeeper-GPU, and the encoding actor plays a critical role in efficiency. Since GateKeeper-GPU is an intermediate tool, choosing the right encoding actor in different scenarios can lead to the optimum results. 

\begin{figure}[!htbp]
\centering
\includegraphics[width=0.45\textwidth]{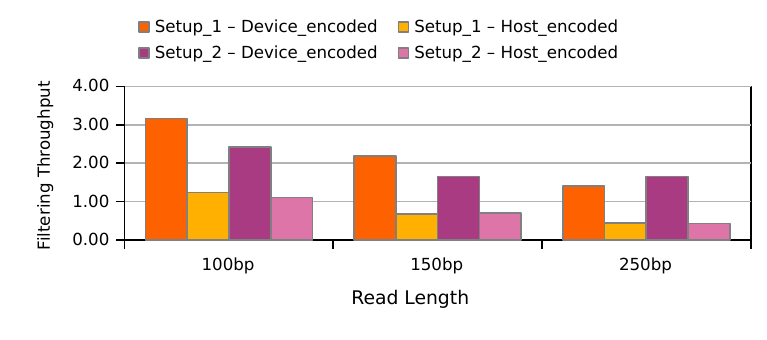}
\caption{\scriptsize Effect of read length on single-GPU GateKeeper-GPU's filtering throughput (millions of filtrations per second) with error threshold \textit{e} = 4. Filtering throughput is calculated with respect to filter time.}
\label{fig:thr_readlength}
\end{figure}

We previously stated that the error threshold has a negligible effect on GateKeeper-GPU's performance and can yield a constant efficiency with increasing error threshold. Apart from the encoding actor, the read length is another factor that influences GateKeeper-GPU's filtering throughput. In longer sequences, GateKeeper-GPU tends to filter with a lower throughput rate, as shown in Figure \ref{fig:thr_readlength}. 

\begin{figure}[!htbp]
\centering
\includegraphics[width=0.45\textwidth]{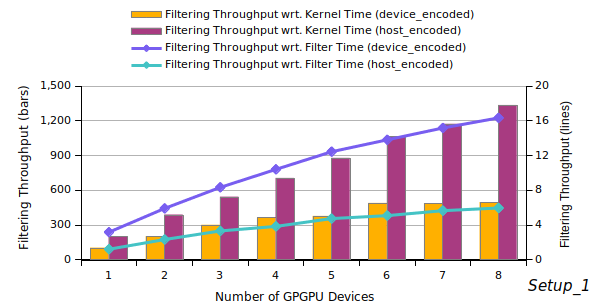}
\caption{\scriptsize Multi-GPU filtering throughput~(millions of filtrations per second) of GateKeeper-GPU in Setup\_1 100bp reads with error threshold \textit{e} = 2. Filtering throughput is calculated with respect to kernel time: bars, and filter time: lines.}
\label{fig:thr_multigpu}
\end{figure}

To understand how GateKeeper-GPU's performance scales with increasing the number of GPGPU devices, we performed tests with 8 GPUs~(Figure \ref{fig:thr_multigpu} and Sup. Figure S.15). With respect to filter time~(lines in Figure \ref{fig:thr_multigpu}), GateKeeper-GPU with encoding in the device experiences a steeper performance increase as the number of devices increases. On the other hand, regarding the kernel time~(bars in Figure \ref{fig:thr_multigpu}), encoding in the device makes GateKeeper-GPU show a slower performance growth with more devices when compared to encoding in the host. Considering the kernel time, adding one more device to the system almost doubles GateKeeper-GPU's throughput in some cases when encoding is done in the host. Therefore, GateKeeper-GPU is better adapted to a multi-GPU environment in a host-encoded fashion. 

In general, GateKeeper-GPU's throughput is lower in Setup\_2 than in Setup\_1 using a single GPU. Setup\_2 does not support data prefetching and has a smaller global memory, both of which are crucial for GateKeeper-GPU's performance, therefore it tends to produce fewer filtrations than Setup\_1.

Due to platform differences, we believe it would not be completely fair to make a comparison between GateKeeper-GPU and other filtering tools in terms of filtering throughput. To create a point of reference, we can briefly express the following observations. Within 40 minutes, GateKeeper-FPGA and SHD can filter up to $\sim$4 trillion and $86$ billion potential mappings, respectively. Regarding kernel time, GateKeeper-GPU can filter more than 7 trillion pairs with 8-GPUs in host-encoding fashion, in Setup\_1 on sequence length 100bp with error threshold 0. However, when considering filter time, the highest number of filtration pairs is $\sim$40 billion, on the same dataset conditions with a device-encoding option on 8 GPUs. 

Considering all of the observations on the filtering throughput of GateKeeper-GPU, we conclude that when the other variables are kept constant, read length and encoding actor (host or device) are critical variables that directly affect the performance, whereas error threshold has a negligible effect. Data prefetching also plays an important role as a platform-dependent factor.   

\subsection{Whole Genome Accuracy \& Performance Analysis}
We evaluate the accuracy and performance of GateKeeper-GPU in full read mapping workflow using four real and two simulated data sets~(Sup. Table S.1).

\begin{table}[!htbp]
\caption{Whole genome mapping information with pre-alignment filtering on real dataset}
\label{tab:whole_gen}
\begin{adjustbox}{max width=0.49\textwidth}
\begin{tabular}{rr|r|r|r|r}
\multicolumn{1}{r}{\textit{mrFAST w/} } & \multicolumn{1}{r}{\textit{-e }} & \multicolumn{1}{r}{\textit{Mappings} } & \multicolumn{1}{r}{\textit{Mapped Reads }} & \multicolumn{1}{r}{\textit{Verification Pairs }} & \multicolumn{1}{c}{\textit{Rejected Pairs (Reduction) }}  \\ 
\hline
\multirow{2}{*}{No Filter}       \rule{0pt}{2ex}        & 0                               & 13,800,412                             & 3,052,036                                  & 257,927,779                                      & NA                                                        \\
                                        & 5                               & 639,841,922                            & 3,887,943                                  & 45,664,847,515                                   & NA                                                        \\ 
\hline
\multirow{2}{*}{GateKeeper-GPU}    \rule{0pt}{2ex}      & 0                               & 13,800,412                             & 3,052,036                                  & 13,824,296                                       & 244,103,483 ( 94 \% )                                     \\
                                        & 5                               & 639,820,825                            & 3,887,939                                  & 4,289,442,302                                    & 41,375,405,213 ( 90 \% )                                  \\
\hline
\end{tabular}
\end{adjustbox}
\begin{minipage}{0.49\textwidth}
\vspace*{0.2cm}
\scriptsize
Mapping information by running mrFAST on ERR240727\_1~(100bp) data set with error thresholds e = 0 and e = 5. The entries represent the number of the corresponding metrics. \textit{-e}: mrFAST's edit distance threshold and error threshold for filtering.
\end{minipage}
\end{table}

For the accuracy tests, we carry out experiments with error thresholds \textit{e} = 0 and \textit{e} = 5 on ERR240727\_1; with error thresholds \textit{e} = 15 for sim\_set\_1 and \textit{e} = 8 for sim\_set\_2. Table \ref{tab:whole_gen}, Sup. Tables  S.24 to S.26 contain mapping information of mrFAST with GateKeeper-GPU. We observe that GateKeeper-GPU successfully filters out faulty mappings and reduces the number of potential mappings correctly by 94\% and 90\% for \textit{e} = 0 and \textit{e} = 5, respectively. We notice a small discrepancy between the total number of mappings and mapped reads produced by mrFAST with and without GateKeeper-GPU when \textit{e} = 5 on the real dataset, as shown in Table \ref{tab:whole_gen}. To analyze these potential false rejects more deeply, we collected some information about the candidate pairs that GateKeeper-GPU rejected and the verification accepted with random sampling. We ran Edlib's global alignment on these samples and found that all these pairs exceeded the error threshold, supporting GateKeeper-GPU's decision to reject. Furthermore, we never experience false rejects in our deep accuracy analysis, as explained in Section 5.1. Therefore, GateKeeper-GPU can \textit{increase} the accuracy of the alignment of mrFAST by rejecting these mappings.

\begin{table}[!htbp]
\caption{Theoretical speedup vs. achieved speedup in verification.}
\label{tab:theoreticalvsachieved}
\begin{adjustbox}{max width=0.49\textwidth}
\begin{tabular}{r|rr|rr}
\multicolumn{1}{r}{}                     & \multicolumn{2}{c}{\textbf{Theoretical DP Time / Speedup} }                     & \multicolumn{2}{c}{\textbf{Achieved DP Time / Speedup} }  \\
\multicolumn{1}{r}{ \textit{mrFAST w/} } & \textit{Setup\_1}                      & \multicolumn{1}{r}{\textit{Setup\_2} } & \textit{Setup\_1}    & \textit{Setup\_2}                  \\ 
\hline  \rule{0pt}{2ex}
No Filter                                & NA                                     & NA                                     & 3.99h                & 4.66h                              \\ 
\hline  \rule{0pt}{2ex}
GateKeeper-GPU (d)                       & \multirow{2}{*}{0.37h / 10.6$\times$ } & \multirow{2}{*}{0.44h / 10.6$\times$ } & 1.08h / 3.7$\times$  & 1.22h / 3.8$\times$                \\
GateKeeper-GPU (h)                       &                                        &                                        & 1.10h / 3.6$\times$  & 1.24h / 3.7$\times$                \\
\hline
\end{tabular}
\end{adjustbox}
\begin{minipage}{0.49\textwidth}
\vspace*{0.2cm}
\scriptsize
Calculations and measurements represent the values for running mrFAST with single-GPU GateKeeper-GPU (encoding in d: device, h: host) on 100bp data set with error threshold \textit{e} = 5. GateKeeper-GPU provides a 90\% reduction in the number of potential mappings. DP: dynamic programming based verification.
\end{minipage}
\end{table}

With respect to the amount of reduction GateKeeper-GPU provides in the number of candidate mappings that enter the verification stage, we calculate the theoretical verification time and speedup by direct proportion. Table \ref{tab:theoreticalvsachieved} compares the theoretical and achieved speedup for the verification stage with GateKeeper-GPU on 100bp data set with \textit{e} = 5. Based on our calculations, we expect a $10.6\times$ speedup on verification, and in practice, verification can achieve up to $3.8\times$ speedup. 

For evaluating how much impact the reduction in the number of candidate mappings that enter verification has on whole genome alignment speedup on a large scale, we construct Table \ref{tab:wholegenspeedup}, Sup. Tables S.24 and S.25. We sum up the time spent on the filtering and verification compared to the verification time of mrFAST when no pre-alignment filtering is used. For filtering time, we consider the kernel time for GateKeeper-GPU. We report only the speedup values of GateKeeper-FPGA since it requires a different platform and the time measurements cannot be scaled for our setting. GateKeeper-GPU has better accuracy than GateKeeper-FPGA and can filter out more candidate mappings while GateKeeper-FPGA performs better than GateKeeper-GPU in terms of accelerating the whole alignment process.

\begin{table}[!htbp]
\caption{Speedup comparison of mrFAST with different pre-alignment filters on a real dataset.}
\label{tab:wholegenspeedup}
\begin{adjustbox}{max width=0.49\textwidth}
\Large
\begin{tabular}{r|ll|ll}
\multicolumn{1}{c}{} & \multicolumn{2}{c}{\textbf{Filtering + DP Time / Speedup} }   & \multicolumn{2}{c}{\textbf{Overall Time / Speedup} }  \\
\rule{0pt}{3ex}
 \textit{mrFAST w/}  & \textit{Setup\_1}    & \multicolumn{1}{l}{\textit{Setup\_2} } & \textit{Setup\_1}    & \textit{Setup\_2}              \\ 
\hline \rule{0pt}{2ex}
No Filter            & 3.99h / NA            & 4.66h / NA                             & 6.85h / NA           & 7.81h / NA                     \\
GateKeeper-GPU (d)   & 1.38h / 2.9$\times$  & 2.85h / 1.6$\times$                    & 4.79h / 1.4$\times$  & 6.63h / 1.2$\times$            \\
GateKeeper-GPU (h)   & 1.38h / 2.9$\times$  & 2.77h / 1.7$\times$                    & 5.26h / 1.3$\times$  & 6.82h / 1.2$\times$            \\
GateKeeper-FPGA      & \multicolumn{2}{c|}{*41$\times$ }                             & \multicolumn{2}{c}{*9.7$\times$ }                     \\
\hline
\end{tabular}
\end{adjustbox}
\begin{minipage}{0.49\textwidth}
\vspace*{0.2cm}
\scriptsize
Speedup comparison between mrFAST's performance with GateKeeper-GPU with single GPU~(encoding in d: device, h: host) and GateKeeper-FPGA on ERR240727\_1~(100bp) data set with error threshold \textit{e} = 5. DP: verification. *values were retrieved from GateKeeper's manuscript~\cite{Alser2017}.
\end{minipage}
\end{table}

For 100bp reads, GateKeeper-GPU can accelerate the verification stage up to $2.9\times$ and $1.7\times$ with Setup\_1 and Setup\_2 when filtering and verification are combined. We directly observe the benefit of data prefetching on Setup\_1 since it creates a larger speedup when compared to Setup\_2. These speedup values reflect on the overall speedup as up to $1.4\times$ and $1.2\times$, respectively.

Even though GateKeeper-GPU can correctly discard $97\%$ of candidate mappings~(Sup. Table S.24), we notice that the kernel time is longer for 300bp and additional operations, such as preparing buffers and data transfer, which are inserted into mrFAST's workflow and required by pre-alignment filtering with GPU, dominate over the time gained from sequence alignment. Also, the data set size is small. Because GateKeeper-GPU exhibits a better performance with large batch sizes, the size of the dataset can be another factor for the absence of speedup. On the 150bp simulated dataset, GateKeeper-GPU can achieve speedup for both verification and overall time in Setup\_1~(Sup. Table S.25). Setup\_2 is deprived of data prefetching and has a smaller global memory, thus the acceleration on verification is not sufficient to reflect the impact on overall mapping duration. These different results of the two device setups validate the effect of data prefetching and global memory size on unified memory usage.  

\subsection{Resource Utilization and Power Analysis}

\subsubsection{Resource Utilization}
The smallest number of registers that the GateKeeper-GPU kernel uses to complete its functionalities fully is 40 registers. The register count can increase to 48 registers per thread in different cases. The maximum theoretical occupancy that can be reached with 48 registers per thread is $63\%$, but the number of threads per block should be at most 256, a quarter of the maximum number of threads per block. Using fewer threads decreases the batch size for a single transfer between host and device, eventually increasing the number of transfers. In our experience, the majority of the time is spent on preprocessing rather than on the kernel. Hence, having the smallest number of transfers is optimal in our scenario, and we opt to set the maximum number of threads for maximized batch size. 

Following this logic, GateKeeper-GPU's theoretical warp occupancy is $50\%$.  In Setup\_1 with encoding in device option, the average achieved occupancy is $48.5\%$ and $49.2\%$ for 100bp and 250bp sequence sets, respectively. When the sequences are encoded in the host, the average occupancy becomes $47.5\%$ and $48.9\%$. Likewise, in Setup\_2 with encoding in the device option, the average achieved occupancy is $46.8\%$ and $48.7\%$ for 100bp and 250bp sequence sets. When sequences are encoded in the host, the average occupancy becomes $44.6\%$ and $47.8\%$, respectively. 

GateKeeper-GPU's achieved occupancy is very close to theoretical warp occupancy. This shows that the warp scheduler can issue new instructions with negligible or no stalls, and the theoretical number of warps is almost reached while SM is active. Therefore, within this limit of register count, the workload within and between the blocks is balanced. On the other hand, because the occupancy is half of its maximum value, average warp execution efficiency is $79.1\%$ and $74.5\%$ in Setup\_1; $80.2\%$ and $76\%$ in Setup\_2 for the 100bp dataset with encoding on device and host, respectively. For the 250bp data set in both device setups, warp execution efficiency is always above $98\%$ on average. This difference created by increasing read length can indicate that occupancy is already at the optimum level for hiding latency in longer sequences.  

GateKeeper-GPU's priority is the effective and full utilization of computing resources to achieve the highest throughput possible. SM efficiency is always above $98\%$ on average and never below $95\%$ regardless of the read length and the encoding actor in both device setups. This high multiprocessor activity shows that the SM(s) are almost never idle during execution. 
\subsubsection{Power Consumption Analysis}
The power consumption of a single GPGPU device for running GateKeeper-GPU is shown in Table \ref{tab:power} and Sup. Table S.27. For 100bp and 250bp datasets, we run with error thresholds 4 and 10, respectively, to evaluate the average energy use of the kernel. The first observation is the negligible effect of the encoding actor on the kernel power consumption when the read length is 100bp. Results show that encoding the sequences on a device or host does not create a huge difference. Even though encoding in the device puts more workload on the device, the energy consumption is almost the same on average for 100bp reads in Setup\_1  owing to effective parallelism~(Table 6). In general, what creates a difference in power consumption is the increase in the read length. The kernel tends to use more power in longer sequences due to increased memory usage and, eventually, more word processing.

\begin{table}[!htbp]
\centering
\caption{Power consumption of GateKeeper-GPU in Setup\_1.}
\label{tab:power}
\begin{adjustbox}{max width=0.49\textwidth}
\begin{tabular}{r|rr|rr}
\multicolumn{1}{c}{}                     & \multicolumn{2}{c}{\textbf{Device-encoded} } & \multicolumn{2}{c}{\textbf{Host-encoded} }  \\
\multicolumn{1}{r}{\textit{Power (mW) }} & \textit{100bp}  & \textit{250bp}             & \textit{100bp}  & \textit{250bp}            \\ 
\hline \rule{0pt}{2ex}
min                                      & 8,901           & 8,702                      & 8,803           & 8,628                     \\
max                                      & 113,218         & 238,701                    & 157,730         & 260,681                   \\
average                                  & 61,868          & 89,023                     & 61,881          & 77,109                    \\
\hline
\end{tabular}
\end{adjustbox}
\begin{minipage}{0.49\textwidth}
\vspace*{0.2cm}
\scriptsize
Power Consumption~(milliwatt) for single GPU in Setup\_1. The values were obtained by running the CUDA command-line profiler nvprof. 
\end{minipage}
\end{table}

\section{Conclusion}
Having compute-intensive nature and heavy workload of data make the verification stage a bottleneck for the entire read mapping process. Pre-alignment filters are designed to facilitate verification by means of reducing the workload as accurately and fast as possible. Different techniques and hardware platforms are utilized for a quick filtering experience. In that sense, GateKeeper-GPU positions itself at a middle level of being the most accurate and fastest pre-alignment filtering tool. Compared to GateKeeper, which was implemented in FPGA, GateKeeper-GPU is more accurate but its benefit on the acceleration of the entire read mapping process is smaller. On the other hand, being a GPGPU tool makes it preferable to an FPGA tool. Since it is implemented on GPU, it is also much more promising for further improvements. 

We believe that, although GateKeeper-GPU still brings benefits, it is more advantageous to consider it when a new short read alignment tool is constructed with a verification-aware design rather than adapting it to an existing read mapping workflow since it requires extra steps for filtration such as encoding. With a well-developed hardware-software co-design of read mapping, it can have a powerful impact on the entire mapping procedure. For that reason, we provide GateKeeper-GPU with two different modes in encoding, both of which can be desirable in different scenarios. If the read mapper is designed to process encoded sequences in its original workflow, GateKeeper-GPU's encoding can be skipped, and the host-encoded version can be utilized. In other scenarios where the time spent for encoding can be overlapped with other operations and be hidden during kernel execution, a device-encoded version can be useful. 

The actual reasons that degrade GateKeeper-GPU's overall performance are tied to preprocessing work that has to be carried out for kernel execution. In future work, we intend to solve these problems and improve the extra time spent on the preprocessing to bring out GateKeeper-GPU's best performance. In addition, we notice that GateKeeper-GPU mainly utilizes L2 cache with an average hit rate of $86.2\%$ rather than unified/texture L1 cache. The hit rate of unified/texture L1 cache is $31.2\%$ on average, which is low. Improving cache utilization is another aim for enabling more efficient kernel activity. 

GateKeeper-GPU brings many benefits over its original work, even though the acceleration is less. It also has comparable results with similar tools. Having the issues resolved, the improvements can be more pronounced. Therefore, GateKeeper-GPU is a promising pre-alignment tool with a wide range of platform support owing to its simple design and GPGPU codebase.

\ifCLASSOPTIONcompsoc
  \section*{Acknowledgments}
\else
  \section*{Acknowledgment}
\fi

This work was partially supported by an EMBO Installation Grant~(IG-2521) to CA. The accuracy analysis for Table S.26 reported in this paper was partially performed at TUBITAK ULAKBIM, High Performance and Grid Computing Center (TRUBA resources). We thank Ricardo Román-Brenes for the assistance with illustrations and helpful comments. An extended abstract of this work was previously presented in HiCOMB~(IEEE IPDPSW-doi:\href{https://doi.org/10.1109/IPDPSW52791.2021.00039}{10.1109/IPDPSW52791.2021.00039}) in 2021.

\ifCLASSOPTIONcaptionsoff
  \newpage
\fi



\bibliographystyle{IEEEtran}
\bibliography{bibtex}
%





\section*{Supplementary material}

# Supplementary Materials

## I. Background

```
Error Threshold    = 2                                        0
Reference Segment  = A C C T G C C A C C C G G A G T C
Read               = T C G A G A T T A A A T C T C C T

Bitvectors after binary encoding:
    Reference S.   = 00 01 01 11 10 01 01 00 01 01 01 10 10 00 10 11 01
    Read           = 11 01 10 00 00 00 11 11 00 00 00 10 01 11 01 01 11
```

```
Hamming mask and amended mask for E = 0:                      1
    H[0]    = 11 00 11 11 11 11 01 01 11 01 01 00 11 11 11 10 10
    A[0]    = 1  1  1  1  1  1  1  1  1  1  1  1  1  1  1  1  1
            e = 17
            e > E, therefore ASM begins.
```

```
Deletion mask (D1) and Insertion mask (I1) their amended versions for E = 1:   2
    Ref         =       00 01 01 11 10 01 01 00 01 01 01 10 10 00 10 11 01
    S.Read (D1) = >>        11 01 10 00 00 00 11 11 00 00 00 10 01 11 01 01 11
    S.Read (I1) = <<     11 01 10 00 00 00 11 11 00 00 00 10 01 11 01 01 11

    H[D1]    =              00 01 11 10 01 01 00 01 01 01 10 10 00 00
    A[D1]    =        0  1  1  1  1  1  1  1  1  1  1  1  1  1  1  0  0

    H[I1]    =           00 01 11 10 10 10 00 01 01 11 01 11 00 00
    A[I1]    =        0  1  1  1  1  1  1  1  1  1  1  1  1  1  1  0  0
```

```
Deletion mask (D2) and Insertion mask (I2) their amended versions for E = 2:   3
    Ref         =       00 01 01 11 10 01 01 00 01 01 01 10 10 00 10 11 01
    S.Read (D2) = >>        11 01 10 00 00 00 11 11 00 00 00 10 01 11 01 01 11
    S.Read (I2) = <<     11 01 10 00 00 00 11 11 00 00 00 10 01 11 01 01 11

    H[D2]    =              00 00 10 00 01 10 10 00 00 11 00 00 00 00
    A[D2]    =        0  0  1  1  1  1  1  1  1  1  1  1  1  0  0  0  0

    H[I2]    =           00 00 11 01 10 10 10 00 01 01 11 01 11 00 00
    A[I2]    =        0  0  1  1  1  1  1  1  1  1  1  1  1  1  1  0  0
```

```
AND Operation and Windowed Error Counting:                    4
    Reference Segment = ..AC CTGC CACC CGGA GTC. .
    Read              = ..TC GAGA TTAA ATCT CCT. .

    A[0]    =  0011 1111 1111 1111 1110 0
    A[D1]   =  0001 1111 1111 1111 1100 0
    A[I1]   =  0011 1111 1111 1111 1100 0
    A[D2]   =  0000 1111 1111 1111 1000 0
    A[I2]   =  0011 1111 1111 1111 1111 0
            &
    AND mask        =  0000 1111 1111 1111 1000 0
    Error counting:    0   1   1   1   1   1   0
```

```
                    e = 4                                     5
Decision:
    Since e > E;
        The read and reference segment pair is REJECTED.
```

```
Blue    : Deletion masks
Red     : Insertion masks
Green   : Amended bits
H[]     : Hamming mask
A[]     : Amended mask
S.Read  : Shifted Read Mask
ASM     : Approximate String Matching
E       : Error Threshold
e       : Number of Errors Found in the Current Step
```

**Figure S. 1:** GateKeeper [1] workflow for error threshold *e = 2*

## II. Methods

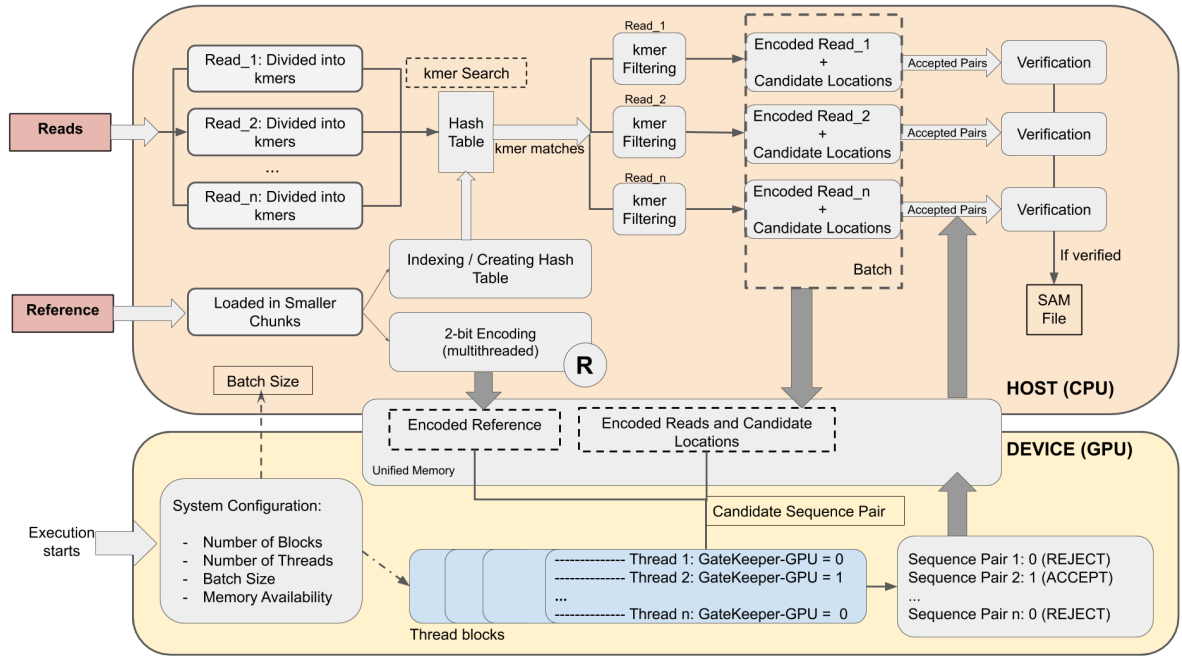

**Figure S. 2:** Workflow of mrFAST with GateKeeper-GPU. Box *R* shows multithreaded encoding the reference and loading it to unified memory. Gray arrows show the data transfers between device and host on unified memory.

## III. Datasets

**Table S. 1:** Details of the Datasets

|  | Number of Reads | Read Length | Undefined Pairs | Details |
|---|---|---|---|---|
| *Whole Genome* |  |  |  |  |
| *sim_set_1 | 100,000 | 300bp | NA | Simulated, rich deletion profile |
| sim_set_2 | 1,000,000 | 150bp | NA | Simulated, low indel profile |
| ERR240727_1 | 4,081,242 | 100bp | NA | Real |
| *Accuracy 5.1.1* |  |  |  |  |
| *Set_3 | 30,000,000 | 100bp | 92,414 | ERR240727_1, mrFAST *-e* = 5 |
| *Set_6 | 30,000,000 | 150bp | 15,141 | SRR826460_1, mrFAST *-e* = 6 |
| *Set_10 | 30,000,000 | 250bp | 379,292 | SRR826471_1, mrFAST *-e* = 12 |
| Minimap2 sets | 30,000,000 | 100bp | Irregular | ERR240727_1, minimap2 *e* = {0, 1, ..., 10} |
| BWA-MEM sets | Irregular | 100bp | Irregular | ERR240727_1, bwa-mem *e* = {0, 1, ..., 10} |
| *Accuracy 5.1.2* |  |  |  |  |
| *Set_1 | 30,000,000 | 100p | 28,009 | ERR240727_1, mrFAST *-e* = 2, *low edit profile* |
| *Set_4 | 30,000,000 | 100p | 31,487 | ERR240727_1, mrFAST *-e* = 40, *high edit profile* |
| *Set_5 | 30,000,000 | 150bp | 30,142 | SRR826460_1, mrFAST *-e* = 4, *low edit profile* |
| *Set_8 | 30,000,000 | 150bp | 309 | SRR826460_1, mrFAST *-e* = 70, *high edit profile* |
| *Set_9 | 30,000,000 | 250bp | 35,072 | SRR826471_1, mrFAST *-e* = 8, *low edit profile* |
| *Set_12 | 30,000,000 | 250bp | 4,763,682 | SRR826471_1, mrFAST -e = 100, *high edit profile* |
| *Filtering Throughput* |  |  |  |  |
| *Set_3 | 30,000,000 | 100bp | 92,414 | ERR240727_1, mrFAST *-e* = 5, *high edit profile* |
| *Set_7 | 30,000,000 | 150bp | 329 | SRR826460_1, mrFAST *-e* = 10, *high edit profile* |
| *Set_11 | 30,000,000 | 250bp | 1,273,260 | SRR826471_1, mrFAST *-e* = 15, *high edit profile* |

Asteriks (*) shows the datasets used in GateKeeper [1].
Undefined pairs show the number of sequence pairs that contain the unknown base call character '*N*'.

## IV. Accuracy of GateKeeper-GPU with respect to Edlib

Table S.2 to Table S.6 show the false accept analysis of GateKeeper-GPU with respect to Edlib's [2] global alignment. The reason why we prefer global alignment results instead of semi-global alignment is that GateKeeper-GPU applies filtering for the pairs which are highly expected to match. For Minimap2 and BWA-MEM data sets, we use the same error threshold for filtering, that is set in the mapper to produce candidate pair, as different from mrFAST-generated data sets. We evaluate the accuracy of GateKeeper-GPU by using two metrics: false accept rate and true reject rate. In *false accept rate*, we report the percentage of number of falsely accepted pairs by GateKeeper-GPU over the number of rejected pairs by Edlib. In *true reject rate*, we report the percentage of number of correctly rejected pairs by GateKeeper-GPU over the total number of rejected pairs by Edlib. Since GateKeeper-GPU directly accepts the sequence pairs that contain the unknown base call character '*N*' (we denote these pairs as *undefined*), in order to observe the real false accept count, undefined pairs are also made accepted by Edlib in all of the tests in this section. Details of the datasets can be obtained from Table S.1. In all of these tests, GateKeeper-GPU's false reject count is always zero.

**Table S. 2:** False Accept Analysis Table (100bp) of Figure 4

| | Edlib | | GateKeeper-GPU | | | | |
|---|---|---|---|---|---|---|---|
| *Error Threshold* | *Accepted* | *Rejected* | *Accepted* | *Rejected* | *False Accept Count* | *False Accept Rate (%)* | *True Reject Rate (%)* |
| 0 | 104,403 | 29,895,597 | 104,403 | 29,895,597 | 0 | 0.00 | 100.00 |
| 1 | 136,979 | 29,863,021 | 165,125 | 29,834,875 | 28,146 | 0.09 | 99.91 |
| 2 | 201,392 | 29,798,608 | 336,884 | 29,663,116 | 135,492 | 0.45 | 99.55 |
| 3 | 299,313 | 29,700,687 | 719,228 | 29,280,772 | 419,915 | 1.41 | 98.59 |
| 4 | 427,108 | 29,572,892 | 1,589,521 | 28,410,479 | 1,162,413 | 3.93 | 96.07 |
| 5 | 583,032 | 29,416,968 | 3,091,304 | 26,908,696 | 2,508,272 | 8.53 | 91.47 |
| 6 | 767,631 | 29,232,369 | 6,158,981 | 23,841,019 | 5,391,350 | 18.44 | 81.56 |
| 7 | 983,575 | 29,016,425 | 9,392,090 | 20,607,910 | 8,408,515 | 28.98 | 71.02 |
| 8 | 1,243,411 | 28,756,589 | 12,547,826 | 17,452,174 | 11,304,415 | 39.31 | 60.69 |
| 9 | 1,561,830 | 28,438,170 | 15,002,035 | 14,997,965 | 13,440,205 | 47.26 | 52.74 |
| 10 | 1,960,522 | 28,039,478 | 17,210,612 | 12,789,388 | 15,250,090 | 54.39 | 45.61 |

False accept analysis of GateKeeper-GPU with respect to Edlib in Set_3. The dataset contains mrFAST's candidate pairs. Undefined pairs (92,414) are included in accepted pairs for both Edlib and GateKeeper-GPU.

**Table S. 3:** False Accept Analysis Table (150bp) of Figure S.3

| | Edlib | | GateKeeper-GPU | | | | |
|---|---|---|---|---|---|---|---|
| *Error Threshold* | *Accepted* | *Rejected* | *Accepted* | *Rejected* | *False Accept Count* | *False Accept Rate (%)* | *True Reject Rate (%)* |
| 0 | 264,061 | 29,735,939 | 264,061 | 29,735,939 | 0 | 0.00 | 100.00 |
| 1 | 339,183 | 29,660,817 | 365,369 | 29,634,631 | 26,186 | 0.09 | 99.91 |
| 3 | 496,836 | 29,503,164 | 832,921 | 29,167,079 | 336,085 | 1.14 | 98.86 |
| 4 | 627,847 | 29,372,153 | 1,391,221 | 28,608,779 | 763,374 | 2.60 | 97.40 |
| 6 | 1,006,666 | 28,993,334 | 3,705,826 | 26,294,174 | 2,699,160 | 9.31 | 90.69 |
| 7 | 1,241,736 | 28,758,264 | 5,837,388 | 24,162,612 | 4,595,652 | 15.98 | 84.02 |
| 9 | 1,755,063 | 28,244,937 | 12,940,109 | 17,059,891 | 11,185,046 | 39.60 | 60.40 |
| 10 | 2,024,813 | 27,975,187 | 16,172,680 | 13,827,320 | 14,147,867 | 50.57 | 49.43 |
| 12 | 2,606,248 | 27,393,752 | 20,339,185 | 9,660,815 | 17,732,937 | 64.73 | 35.27 |
| 13 | 2,938,643 | 27,061,357 | 21,536,343 | 8,463,657 | 18,597,700 | 68.72 | 31.28 |
| 15 | 3,745,005 | 26,254,995 | 23,563,237 | 6,436,763 | 19,818,232 | 75.48 | 24.52 |

False accept analysis of GateKeeper-GPU with respect to Edlib in Set_6. The dataset contains mrFAST's candidate pairs. Undefined pairs (15,141) are included in the accepted pairs for both Edlib and GateKeeper-GPU.

**Table S. 4:** False Accept Analysis Table (250bp) of Figure S.4

| | Edlib | | GateKeeper-GPU | | | | |
|---|---|---|---|---|---|---|---|
| *Error Threshold* | *Accepted* | *Rejected* | *Accepted* | *Rejected* | *False Accept Count* | *False Accept Rate (%)* | *True Reject Rate (%)* |
| 0 | 422,817 | 29,577,183 | 422,827 | 29,577,173 | 10 | 0.00 | 100.00 |
| 2 | 467,342 | 29,532,658 | 479,872 | 29,520,128 | 12,530 | 0.04 | 99.96 |
| 5 | 498,223 | 29,501,777 | 587,918 | 29,412,082 | 89,695 | 0.30 | 99.70 |
| 7 | 524,364 | 29,475,636 | 793,255 | 29,206,745 | 268,891 | 0.91 | 99.09 |
| 10 | 584,475 | 29,415,525 | 1,581,379 | 28,418,621 | 996,904 | 3.39 | 96.61 |
| 12 | 636,191 | 29,363,809 | 3,464,025 | 26,535,975 | 2,827,834 | 9.63 | 90.37 |
| 15 | 725,455 | 29,274,545 | 9,177,446 | 20,822,554 | 8,451,991 | 28.87 | 71.13 |
| 17 | 788,472 | 29,211,528 | 13,113,437 | 16,886,563 | 12,324,965 | 42.19 | 57.81 |
| 20 | 885,488 | 29,114,512 | 17,887,612 | 12,112,388 | 17,002,124 | 58.40 | 41.60 |
| 22 | 950,913 | 29,049,087 | 21,548,632 | 8,451,368 | 20,597,719 | 70.91 | 29.09 |
| 25 | 1,051,100 | 28,948,900 | 26,581,310 | 3,418,690 | 25,530,210 | 88.19 | 11.81 |

False accept analysis of GateKeeper-GPU with respect to Edlib in Set_10. The dataset contains mrFAST's candidate pairs. Undefined pairs (379,262) are included in the accepted pairs for both Edlib and GateKeeper-GPU.

**Table S. 5:** False Accept Analysis Table of Figure S.5 for Minimap2 sets

| | Edlib | | GateKeeper-GPU | | | | |
|---|---|---|---|---|---|---|---|
| _Error Threshold_ | _Accepted_ | _Rejected_ | _Accepted_ | _Rejected_ | _False Accept Count_ | _False Accept Rate (%)_ | _True Reject Rate (%)_ |
| 0 | 813,961 | 29,186,039 | 813,961 | 29,186,039 | 0 | 0.00 | 100.00 |
| 1 | 1,020,992 | 28,979,008 | 1,080,900 | 28,919,100 | 60,351 | 0.21 | 99.79 |
| 2 | 1,167,931 | 28,832,069 | 1,333,047 | 28,666,953 | 165,328 | 0.57 | 99.43 |
| 3 | 1,309,257 | 28,690,743 | 1,709,395 | 28,290,605 | 398,621 | 1.39 | 98.61 |
| 4 | 1,463,733 | 28,536,267 | 2,333,802 | 27,666,198 | 869,491 | 3.05 | 96.95 |
| 5 | 1,638,308 | 28,361,692 | 3,346,596 | 26,653,404 | 1,705,665 | 6.01 | 93.98 |
| 6 | 1,836,191 | 28,163,809 | 4,844,286 | 25,155,714 | 2,999,806 | 10.65 | 89.32 |
| 7 | 2,057,082 | 27,942,918 | 6,703,603 | 23,296,397 | 4,635,784 | 16.59 | 83.37 |
| 8 | 2,303,071 | 27,696,929 | 9,005,034 | 20,994,966 | 6,682,829 | 24.13 | 75.80 |
| 9 | 2,572,631 | 27,427,369 | 11,382,399 | 18,617,601 | 8,784,715 | 32.03 | 67.88 |
| 10 | 2,873,114 | 27,126,886 | 14,001,878 | 15,998,122 | 11,090,571 | 40.88 | 58.98 |

False accept analysis of GateKeeper-GPU with respect to Edlib. Undefined pairs (26,759) are included in the accepted pairs for both Edlib and GateKeeper-GPU.

**Table S. 6:** False Accept Analysis Table of Figure S.6 for BWA-MEM sets

| | | Edlib | | GateKeeper-GPU | | | | |
|---|---|---|---|---|---|---|---|---|
| _Error Threshold_ | _Dataset Size_ | _Accepted_ | _Rejected_ | _Accepted_ | _Rejected_ | _False Accept Count_ | _False Accept Rate (%)_ | _True Reject Rate (%)_ |
| 0 | 17,725 | 12,298 | 5,427 | 12,298 | 5,427 | 0 | 0.00 | 100.00 |
| 1 | 1,692 | 17 | 1,675 | 50 | 1,642 | 33 | 1.97 | 98.03 |
| 2 | 1,723 | 17 | 1,706 | 424 | 1,299 | 407 | 23.86 | 76.14 |
| 3 | 1,721 | 617 | 1,104 | 1,041 | 680 | 424 | 38.41 | 61.59 |
| 4 | 8,904 | 5,223 | 3,681 | 7,219 | 1,685 | 1,996 | 54.22 | 45.78 |
| 5 | 1,688 | 1,196 | 492 | 1,580 | 108 | 384 | 78.05 | 21.95 |
| 6 | 1,681 | 1,339 | 342 | 1,648 | 33 | 309 | 90.35 | 9.65 |
| 7 | 1,674 | 1,457 | 217 | 1,668 | 6 | 211 | 97.24 | 2.76 |
| 8 | 1,677 | 1,539 | 138 | 1,676 | 1 | 137 | 99.28 | 0.72 |
| 9 | 1,681 | 1,578 | 103 | 1,681 | 0 | 103 | 100.00 | 0.00 |
| 10 | 1,665 | 1,597 | 68 | 1,665 | 0 | 68 | 100.00 | 0.00 |

False accept analysis of GateKeeper-GPU with respect to Edlib in BWA-MEM sets. Undefined pairs are included in the number of accepted pairs for both Edlib and GateKeeper-GPU.

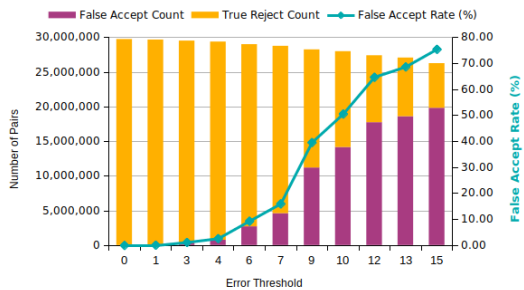

**Figure S. 3:** False Accept Analysis - 150bp

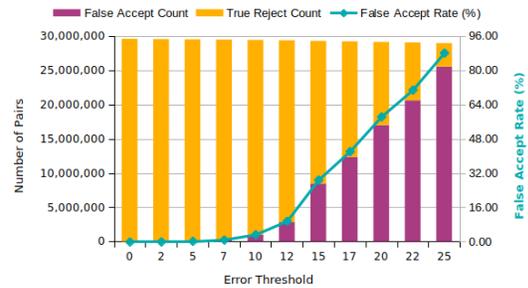

**Figure S. 4:** False Accept Analysis - 250bp

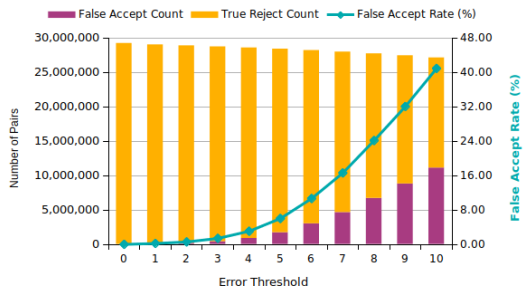

**Figure S. 5:** False accept analysis for Minimap2 sets

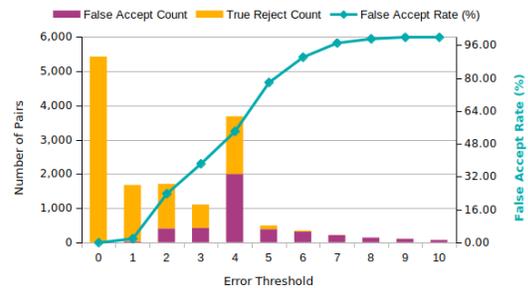

**Figure S. 6:** False accept analysis for BWA-MEM sets

## V. Comparison with Other Pre-alignment Filters

Table S.7 to Table S.12 show the false accept comparison of GateKeeper-GPU against GateKeeper [1] (denoted as *GateKeeper-FPGA*), SHD [3], MAGNET [4], Shouji [5] and SneakySnake [6] with respect to Edlib's [2] global alignment. The results are also represented by Figure 5, and Figure S.7 to Figure S.11. Since the other filters do not distinguish the sequence pairs that contain the unknown base call character '*N*' (denoted as *undefined*), in order to make a fair comparison, we include undefined pairs in GateKeeper-GPU's false accept count.

**Table S. 7:** False Accept Comparison Table (100bp) of Figure 5

| Error Threshold | GateKeeper-GPU | GateKeeper-FPGA | SHD | Shouji | MAGNET | SneakySnake |
|---|---|---|---|---|---|---|
| 0 | 28,009 (0) | 0 | 0 | 0 | 963,941 | 0 |
| 1 | 672,164 (644,200) | 783,185 | 783,185 | 333,320 | 800,099 | 12,473 |
| 2 | 2,290,693 (2,263,036) | 2,704,128 | 2,704,128 | 1,283,004 | 1,876,518 | 77,165 |
| 3 | 4,324,420 (4,297,528) | 5,237,529 | 5,237,529 | 2,674,876 | 2,428,301 | 234,003 |
| 4 | 6,744,070 (6,718,357) | 8,231,507 | 8,231,507 | 4,399,886 | 2,662,902 | 484,179 |
| 5 | 9,354,269 (9,330,017) | 11,195,124 | 11,195,124 | 6,452,280 | 2,916,838 | 795,582 |
| 6 | 12,092,022 (12,069,082) | 13,781,651 | 13,781,651 | 9,373,309 | 3,406,303 | 1,240,276 |
| 7 | 13,085,652 (13,064,066) | 14,283,519 | 14,283,519 | 11,113,616 | 4,026,433 | 1,815,478 |
| 8 | 13,139,626 (13,119,339) | 13,814,295 | 13,814,295 | 11,990,529 | 4,745,672 | 2,567,290 |
| 9 | 12,264,194 (12,245,362) | 13,105,305 | 13,105,305 | 11,693,396 | 5,319,627 | 3,331,944 |
| 10 | 10,929,703 (10,912,255) | 11,389,103 | 11,389,103 | 10,664,722 | 5,673,172 | 4,020,164 |

False accept comparison between pre-alignment tools for Set_1. GateKeeper-GPU values show number of false accepts including undefined pairs outside of parenthesis and excluding undefined pairs inside of parenthesis. Figure 5 was drawn with false accept counts including undefined pairs that are outside of parenthesis. The values for other filters were retrieved from Shouji's supplementary material.

**Table S. 8:** False Accept Comparison Table (100bp) of Figure S.7

| Error Threshold | GateKeeper-GPU | GateKeeper-FPGA | SHD | Shouji | MAGNET | SneakySnake |
|---|---|---|---|---|---|---|
| 0 | 31,487 (0) | 0 | 0 | 0 | 7 | 0 |
| 1 | 31,501 (14) | 14 | 14 | 2 | 5 | 0 |
| 2 | 31,767 (280) | 155 | 155 | 15 | 2 | 0 |
| 3 | 32,689 (1,202) | 1,196 | 1,196 | 216 | 4 | 1 |
| 4 | 40,692 (9,205) | 7,436 | 7,436 | 1,986 | 13 | 3 |
| 5 | 71,158 (39,671) | 32,792 | 32,792 | 10,551 | 82 | 13 |
| 6 | 193,539 (162,052) | 155,134 | 155,134 | 57,258 | 298 | 69 |
| 7 | 435,611 (404,124) | 417,444 | 417,444 | 214,005 | 1,030 | 289 |
| 8 | 951,114 (919,627) | 1,031,480 | 1,031,480 | 675,029 | 3,129 | 1,081 |
| 9 | 1,943,019 (1,911,532) | 29,997,022 | 29,997,022 | 1,742,476 | 8,234 | 3,563 |
| 10 | 3,710,604 (3,679,117) | 29,998,373 | 29,998,373 | 3,902,535 | 19,013 | 9,698 |

False accept comparison between pre-alignment tools for Set_40. GateKeeper-GPU values show number of false accepts including undefined pairs outside of parenthesis and excluding undefined pairs inside of parenthesis. Figure S.7 was drawn with false accept counts including undefined pairs that are outside of parenthesis. The values for other filters were retrieved from Shouji's supplementary material.

**Table S. 9:** False Accept Comparison Table (150bp) of Figure S.8

| Error Threshold | GateKeeper-GPU | GateKeeper-FPGA | SHD | Shouji | MAGNET |
|---|---|---|---|---|---|
| 0 | 30,142 (0) | 0 | 0 | 0 | 428,412 |
| 1 | 171,256 (141,961) | 173,573 | 173,573 | 113,519 | 156,891 |
| 3 | 1,632,544 (1,603,900) | 2,080,279 | 2,080,279 | 1,539,365 | 725,873 |
| 4 | 3,118,355 (3,090,152) | 4,023,762 | 4,023,762 | 3,042,831 | 1,064,344 |
| 6 | 6,681,929 (6,654,933) | 9,258,602 | 9,258,602 | 6,025,592 | 1,430,272 |
| 7 | 9,016,979 (8,990,561) | 12,481,853 | 12,481,853 | 8,219,336 | 1,532,024 |
| 9 | 15,109,160 (15,083,838) | 22,076,837 | 22,076,837 | 14,568,337 | 1,874,734 |
| 10 | 17,023,658 (16,998,826) | 21,341,979 | 21,341,979 | 16,920,389 | 2,194,275 |
| 12 | 18,335,496 (18,311,754) | 19,868,151 | 19,868,151 | 18,270,597 | 3,294,672 |
| 13 | 18,145,432 (18,122,415) | 19,082,528 | 19,082,528 | 18,095,207 | 4,066,617 |
| 15 | 16,953,324 (16,932,083) | 17,353,835 | 17,353,835 | 16,993,568 | 5,810,797 |

False accept comparison between pre-alignment tools for Set_5. GateKeeper-GPU values show number of false accepts including undefined pairs outside of parenthesis and excluding undefined pairs inside of parenthesis. Figure S.8 was drawn with false accept counts including undefined pairs that are outside of parenthesis. The values for other filters were retrieved from Shouji's supplementary material.

**Table S. 10:** False Accept Comparison Table (150bp) of Figure S.9

| Error Threshold | GateKeeper-GPU | GateKeeper-FPGA | SHD | Shouji | MAGNET |
|---:|---:|---:|---:|---:|---:|
| 0 | 309 (0) | 0 | 0 | 0 | 126 |
| 1 | 365 (58) | 58 | 58 | 43 | 42 |
| 3 | 407 (100) | 90 | 90 | 83 | 35 |
| 4 | 573 (266) | 267 | 267 | 137 | 28 |
| 6 | 13,606 (13,299) | 18,110 | 18,110 | 6,259 | 25 |
| 7 | 64,840 (64,533) | 79,418 | 79,418 | 27,092 | 27 |
| 9 | 564,241 (563,934) | 29,698,666 | 29,698,666 | 404,742 | 108 |
| 10 | 1,049,599 (1,049,292) | 29,999,388 | 29,999,388 | 935,486 | 231 |
| 12 | 2,490,712 (2,490,405) | 29,999,290 | 29,999,290 | 2,514,950 | 965 |
| 13 | 3,677,914 (3,677,607) | 29,999,204 | 29,999,204 | 3,693,298 | 2,018 |
| 15 | 7,692,574 (7,692,267) | 29,998,847 | 29,998,847 | 8,034,737 | 8,448 |

False accept comparison between pre-alignment tools for Set_8. GateKeeper-GPU values show the number of false accepts including undefined pairs outside of parenthesis and excluding undefined pairs inside of parenthesis. The figure Figure S.9 was drawn with false accept counts including undefined pairs that are outside of parenthesis. The values for other filters were retrieved from Shouji's supplementary material.

**Table S. 11:** False Accept Comparison Table (250bp) of Figure S.10

| Error Threshold | GateKeeper-GPU | GateKeeper-FPGA | SHD | Shouji | MAGNET | SneakySnake |
|---:|---:|---:|---:|---:|---:|---:|
| 0 | 35,075 (3) | 0 | 0 | 0 | 479,104 | 0 |
| 2 | 250,322 (215,613) | 238,368 | 238,368 | 174,366 | 143,066 | 12,319 |
| 5 | 1,242,873 (1,208,633) | 1,546,126 | 1,546,126 | 1,071,218 | 226,864 | 38,814 |
| 7 | 3,113,200 (3,079,257) | 3,933,916 | 3,933,916 | 2,775,419 | 347,819 | 79,246 |
| 10 | 7,283,863 (7,250,529) | 26,816,729 | 26,816,729 | 6,669,084 | 624,927 | 235,689 |
| 12 | 12,260,108 (12,227,208) | 26,137,224 | 26,137,224 | 11,147,373 | 825,468 | 407,799 |
| 15 | 19,039,913 (19,007,867) | 25,084,654 | 25,084,654 | 18,406,823 | 1,066,633 | 705,904 |
| 17 | 21,308,177 (21,276,706) | 24,449,131 | 24,449,131 | 20,971,826 | 1,235,999 | 914,730 |
| 20 | 22,311,079 (22,280,323) | 23,595,168 | 23,595,168 | 22,223,170 | 1,695,351 | 1,364,891 |
| 22 | 22,311,569 (22,281,259) | 23,040,384 | 23,040,384 | 22,271,215 | 2,241,984 | 1,879,428 |
| 25 | 21,843,548 (21,813,824) | 22,142,250 | 22,142,250 | 21,849,454 | 3,514,515 | 3,134,474 |

False accept comparison between pre-alignment tools for Set_9. GateKeeper-GPU values show the number of false accepts including undefined pairs outside of parenthesis and excluding undefined pairs inside of parenthesis. Figure S.10 was drawn with false accept counts including undefined pairs that are outside of parenthesis. The values for other filters were retrieved from Shouji's supplementary material.

**Table S. 12:** False Accept Comparison Table (250bp) of Figure S.11

| Error Threshold | GateKeeper-GPU | GateKeeper-FPGA | SHD | Shouji | MAGNET | SneakySnake |
|---:|---:|---:|---:|---:|---:|---:|
| 0 | 4,763,683 (1) | 0 | 0 | 0 | 53 | 0 |
| 2 | 4,763,696 (49) | 71 | 71 | 55 | 44 | 2 |
| 5 | 4,763,688 (102) | 249 | 249 | 161 | 49 | 6 |
| 7 | 4,763,704 (152) | 698 | 698 | 212 | 48 | 6 |
| 10 | 4,771,455 (7,953) | 29,999,528 | 29,999,528 | 5,627 | 42 | 14 |
| 12 | 4,839,211 (75,739) | 29,999,480 | 29,999,480 | 64,225 | 45 | 22 |
| 15 | 5,481,110 (717,669) | 29,999,425 | 29,999,425 | 775,314 | 82 | 47 |
| 17 | 6,545,084 (1,781,675) | 29,999,377 | 29,999,377 | 2,052,498 | 175 | 106 |
| 20 | 9,894,411 (5,131,063) | 29,999,282 | 29,999,282 | 5,679,869 | 417 | 326 |
| 22 | 14,252,812 (9,489,566) | 29,999,158 | 29,999,158 | 10,277,297 | 593 | 495 |
| 25 | 21,963,183 (17,200,145) | 29,998,867 | 29,998,867 | 19,676,652 | 1,174 | 955 |

False accept comparison between pre-alignment tools for Set_12. GateKeeper-GPU values show the number of false accepts including undefined pairs outside of parenthesis and excluding undefined pairs inside of parenthesis. The figure Figure S.11 was drawn with false accept counts including undefined pairs that are outside of parenthesis. The values for other filters were retrieved from Shouji's supplementary material.

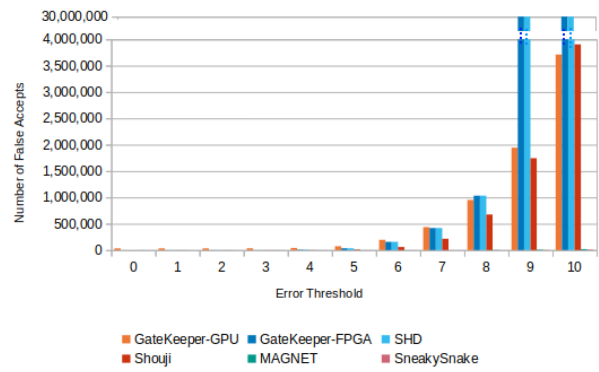

**Figure S. 7:** False accept comparison for high-edit profile in read length 100bp, number of undefined pairs = 31,487

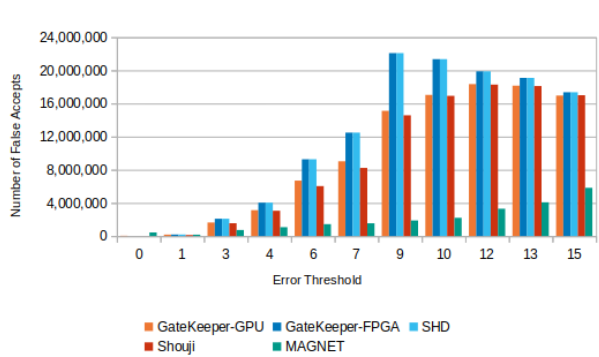

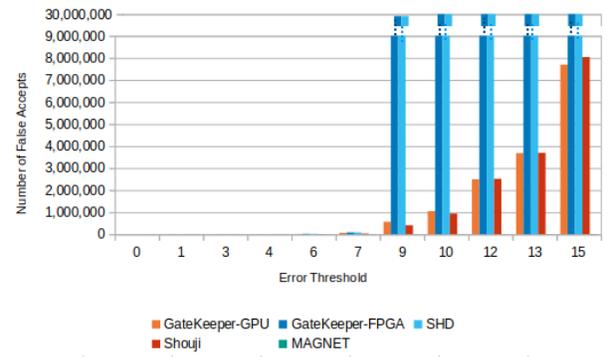

**Figure S. 8:** False accept comparison for low-edit profile in read length 150bp, number of undefined pairs = 30,142

**Figure S. 9:** False accept comparison for high-edit profile in read length 150bp, number of undefined pairs = 309

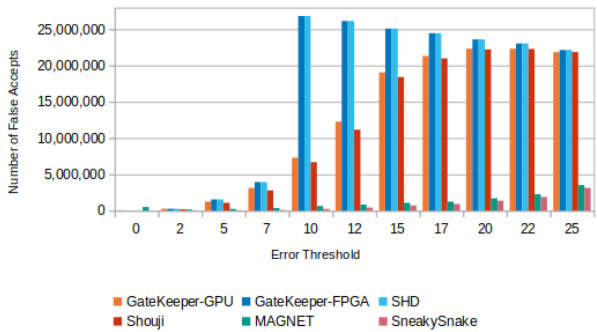

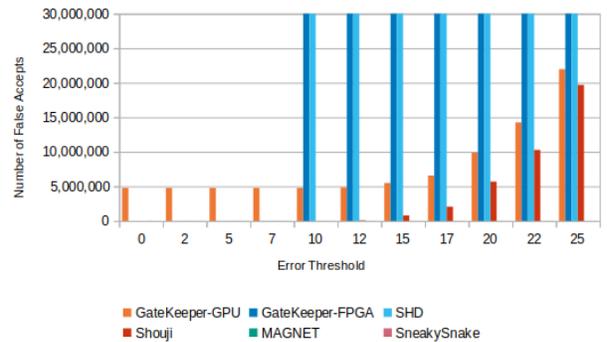

**Figure S. 10:** False accept comparison for low-edit profile in read length 250bp, number of undefined pairs = 35,072

**Figure S. 11:** False Accept Comparison for High-Edit Profile in Read Length 250bp, number of undefined pairs = 4,763,682.

## VI. Filtering Throughput Analysis

For calculating the filtering throughput of GateKeeper-GPU in terms of number of pairs in 40 minutes, we first record the kernel time and filter time taken for filtering of whole data set and report them. Then, we calculate the number of pairs filtered per second and in 40 minutes using these values. Table S.13 to S.15 contain both the unprocessed data for *kernel time* (kt) and *filter time* (ft) in seconds, and finalized results as filtering throughput, which is calculated with respect to kernel time and filter time. Multi-GPU values represent the time of the device that has longest value.

**Table S. 13:** Raw Data of 100bp Kernel and Filter Time for Table 2

|  |  | **GateKeeper-CPU** | | **Device-encoded** | | **Host-encoded** | |
| --- | --- | --- | --- | --- | --- | --- | --- |
|  | *e* | *1-Core* | *12-Cores* | *1-GPU* | *8-GPU* | *1-GPU* | *8-GPU* |
| Setup_1 | kt | 2 | 102.52 | 10.04 | 0.29 | 0.06 | 0.15 | 0.02 |
|  |  | 5 | 194.13 | 18.49 | 0.48 | 0.07 | 0.29 | 0.04 |
|  | ft | 2 | 117.14 | 11.09 | 9.40 | 1.83 | 24.36 | 5.02 |
|  |  | 5 | 204.35 | 19.55 | 9.50 | 1.90 | 24.56 | 5.06 |
| Setup_2 | kt | 2 | 110.70 | 13.21 | 1.75 | NA | 1.00 | NA |
|  |  | 5 | 211.67 | 24.39 | 2.47 | NA | 1.72 | NA |
|  | ft | 2 | 121.47 | 14.73 | 11.74 | NA | 26.49 | NA |
|  |  | 5 | 222.43 | 25.89 | 12.72 | NA | 27.15 | NA |

Kernel time (kt) and filter time (ft) in seconds. *e* = error threshold.

**Table S. 14:** Time and Filtering Throughput of 150bp

|  |  | **GateKeeper-CPU** | | **Device-encoded** | | **Host-encoded** | |
| --- | --- | --- | --- | --- | --- | --- | --- |
|  | *e* | *1-Core* | *12-Cores* | *1-GPU* | *8-GPU* | *1-GPU* | *8-GPU* |
| Setup_1 | kt | 4 | 274.39s - 0.3 | 24.90s - 2.9 | 0.81s - 89.0 | 0.15s - 496.5 | 0.35s - 205.2 | **0.07s - 1,022.0** |
|  |  | 10 | 577.29s - 0.1 | 52.54s - 1.4 | 1.16s - 61.9 | 0.18s - 406.8 | 0.73s - 98.0 | **0.12s - 582.7** |
|  | ft | 4 | 289.96s - 0.2 | 26.37s - 2.7 | 13.74s - 5.2 | **2.82s - 25.5** | 44.14s - 1.6 | 7.73s - 9.3 |
|  |  | 10 | 592.86s - 0.1 | 54.05s - 1.3 | 14.12s - 5.1 | **2.72s - 26.5** | 44.27s - 1.6 | 7.86s - 9.2 |
| Setup_2 | kt | 4 | 262.07s - 0.3 | 28.49s - 2.5 | 3.84s - 18.8 | NA | **2.21s - 32.5** | NA |
|  |  | 10 | 552.53s - 0.1 | 59.80s - 1.2 | 6.01s - 12.0 | NA | **4.36s - 16.5** | NA |
|  | ft | 4 | 278.62s - 0.3 | 30.71s - 2.3 | **18.16s - 4.0** | NA | 43.04s - 1.7 | NA |
|  |  | 10 | 569.17s - 0.1 | 62.00s - 1.2 | **20.41s - 3.5** | NA | 45.23s - 1.6 | NA |

Kernel time (kt) and filter time (ft) in seconds, and the respective filtering throughput that each measurement yields are shown. Filtering throughput is in terms of billions of pairs in 40 minutes, and the bold font indicates the shortest time and highest throughput value in each row. *e* = error threshold.

**Table S. 15:** Time and Filtering Throughput of 250bp

|  |  | **GateKeeper-CPU** | | **Device-encoded** | | **Host-encoded** | |
| --- | --- | --- | --- | --- | --- | --- | --- |
|  | *e* | *1-Core* | *12-Cores* | *1-GPU* | *8-GPU* | *1-GPU* | *8-GPU* |
| Setup_1 | kt | 6 | 575.92s - 0.12 | 53.72s - 1.3 | 1.57s - 45.7 | 0.27s - 271.8 | 0.74s - 97.4 | **0.12s - 611.1** |
|  |  | 10 | 885.18s - 0.08 | 82.17s - 0.9 | 1.87s - 38.4 | 0.29s - 248.6 | 1.17s - 61.8 | **0.22s - 331.4** |
|  | ft | 6 | 601.32s - 0.12 | 56.06s - 1.3 | 21.78s - 3.3 | **4.24s - 17.0** | 69.43s - 1.0 | 11.16s - 6.5 |
|  |  | 10 | 910.18s - 0.08 | 84.54s - 0.9 | 22.06s - 3.3 | **4.11s - 17.5** | 69.97s - 1.0 | 11.33s - 6.4 |
| Setup_2 | kt | 6 | 558.09s - 0.13 | 64.87s - 1.1 | **4.55s - 15.8** | NA | **4.55s - 15.8** | NA |
|  |  | 10 | 862.35s - 0.08 | 98.61s - 0.7 | 6.01s - 12.0 | NA | **6.74s - 10.7** | NA |
|  | ft | 6 | 583.10s - 0.12 | 68.46s - 1.1 | **18.87s - 3.8** | NA | 71.42s - 1.0 | NA |
|  |  | 10 | 887.43s - 0.08 | 102.23s - 0.7 | **20.41s - 3.5** | NA | 73.86s - 1.0 | NA |

Kernel time (kt) and filter time (ft) in seconds, and the respective filtering throughput that each measurement yields are shown. Filtering throughput is in terms of billions of pairs in 40 minutes, and the bold font indicates the shortest time and highest throughput value in each row. *e* = error threshold.

## A. Effect of Error Threshold on Filter Time

Table S.16 contains the data for evaluating the effect of increasing error threshold on the performance of single GPU GateKeeper-GPU and 12-core GateKeeper-CPU. The data is illustrated by Figure S.12.

**Table S. 16:** Effect of Increasing Error Threshold

| $e$ | Setup_1 | | | Setup_2 | | |
|---|---|---|---|---|---|---|
| | 12-core CPU | Device-encoded GPU | Host-encoded GPU | 12-core CPU | Device-encoded GPU | Host-encoded GPU |
| 0 | 12.18 | 22.10 | 73.99 | 14.09 | 17.23 | 69.29 |
| 1 | 21.32 | 23.84 | 68.85 | 25.89 | 16.99 | 68.74 |
| 2 | 28.22 | 22.03 | 68.77 | 34.39 | 17.46 | 69.28 |
| 4 | 41.72 | 21.27 | 69.31 | 51.46 | 18.16 | 70.49 |
| 6 | 56.06 | 21.78 | 69.43 | 68.46 | 18.87 | 71.42 |
| 8 | 70.25 | 21.61 | 69.59 | 85.42 | 20.40 | 72.76 |
| 10 | 84.54 | 22.06 | 69.97 | 102.23 | 20.41 | 73.86 |

Effect of increasing error threshold in multi-core GateKeeper-CPU and single GPU GateKeeper-GPU in terms of filter time (seconds). 250bp pairs were used for the tests. $e$ = error threshold.

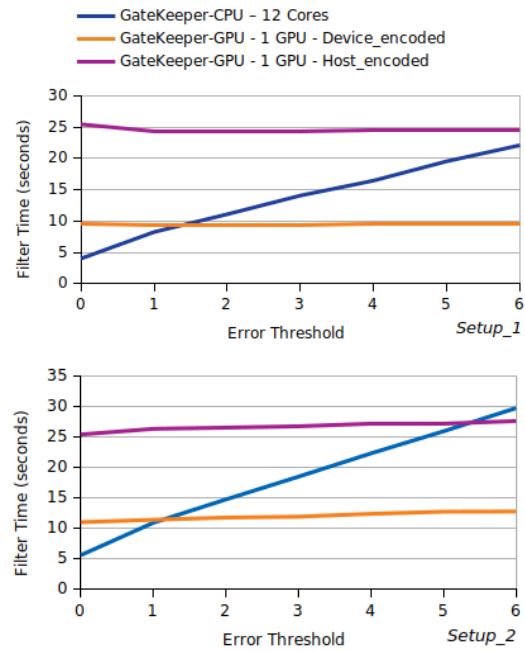

**Figure S. 12:** Effect of increasing error threshold on the performance of GateKeeper-CPU and GateKeeper-GPU by means of filter time (seconds)

## B. Effect of Encoding Actor on Filtering Throughput

Table S.17 to Table S.19 contain the data for evaluating the effect of encoding actor on the performance of GateKeeper-GPU.

**Table S. 17:** Effect of Encoding Actor on Filtering Throughput in 100bp for Figure 6

| | Setup_1 | | | | Setup_2 | | | |
| | Device-encoded | | Host-encoded | | Device-encoded | | Host-encoded | |
| $e$ | Kernel | Filter | Kernel | Filter | Kernel | Filter | Kernel | Filter |
|---|---|---|---|---|---|---|---|---|
| 0 | 110.1 | 3.2 | 699.7 | 1.2 | 25.9 | 2.7 | 83.1 | 1.2 |
| 1 | 113.2 | 3.2 | 282.6 | 1.2 | 19.9 | 2.6 | 39.9 | 1.2 |
| 2 | 102.0 | 3.2 | 198.7 | 1.2 | 17.1 | 2.6 | 30.1 | 1.1 |
| 3 | 91.6 | 3.2 | 149.7 | 1.2 | 15.0 | 2.5 | 24.2 | 1.1 |
| 4 | 72.5 | 3.2 | 122.5 | 1.2 | 13.4 | 2.4 | 20.3 | 1.1 |
| 5 | 62.8 | 3.2 | 103.9 | 1.2 | 12.1 | 2.4 | 17.5 | 1.1 |
| 6 | 57.0 | 3.2 | 89.7 | 1.2 | 11.0 | 2.4 | 15.3 | 1.1 |

Effect of encoding actor (device or host) on filtering throughput of single-GPU GateKeeper-GPU with increasing error threshold. The values represent the number of filtrations in terms of millions per second, produced by GateKeeper-GPU with respect to kernel or filter time. $e$ = error threshold.

**Table S. 18:** Effect of Encoding Actor on Filtering Throughput in 150bp

| | Setup_1 | | | | Setup_2 | | | |
| | Device-encoded | | Host-encoded | | Device-encoded | | Host-encoded | |
| $e$ | Kernel | Filter | Kernel | Filter | Kernel | Filter | Kernel | Filter |
|---|---|---|---|---|---|---|---|---|
| 0 | 28.9 | 2.1 | 537.7 | 0.6 | 11.2 | 1.7 | 56.5 | 0.7 |
| 1 | 32.8 | 2.0 | 193.5 | 0.7 | 10.7 | 1.8 | 26.8 | 0.7 |
| 2 | 35.6 | 2.2 | 138.6 | 0.7 | 9.6 | 1.7 | 20.2 | 0.7 |
| 4 | 37.1 | 2.2 | 85.5 | 0.7 | 7.8 | 1.7 | 13.6 | 0.7 |
| 6 | 34.9 | 2.2 | 62.1 | 0.7 | 6.6 | 1.6 | 10.1 | 0.7 |
| 8 | 30.5 | 2.1 | 48.4 | 0.7 | 5.7 | 1.5 | 8.2 | 0.7 |
| 10 | 25.8 | 2.1 | 40.8 | 0.7 | 5.0 | 1.5 | 6.9 | 0.7 |

Effect of encoding actor (device or host) on filtering throughput of single-GPU GateKeeper-GPU with increasing error threshold. The values represent the number of filtrations in terms of millions per second, produced by GateKeeper-GPU with respect to kernel or filter time. $e$ = error threshold.

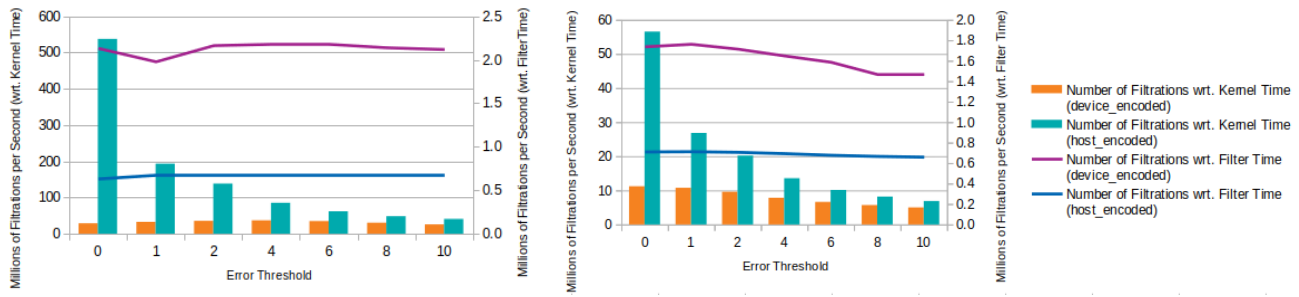

**Figure S. 13:** Effect of encoding actor (device or host) on filtering throughput of single-GPU GateKeeper-GPU with increasing error threshold for read length 150bp in Setup_1 and Setup_2, respectively

**Table S. 19:** Effect of Encoding Actor on Filtering Throughput in 250bp

| | Setup_1 | | | | Setup_2 | | | |
| | Device-encoded | | Host-encoded | | Device-encoded | | Host-encoded | |
| $e$ | Kernel | Filter | Kernel | Filter | Kernel | Filter | Kernel | Filter |
|---|---|---|---|---|---|---|---|---|
| 0 | 15.3 | 1.4 | 328.2 | 0.4 | 11.2 | 1.7 | 36.6 | 0.4 |
| 1 | 16.3 | 1.3 | 132.2 | 0.4 | 10.7 | 1.8 | 17.2 | 0.4 |
| 2 | 17.0 | 1.4 | 92.0 | 0.4 | 9.6 | 1.7 | 12.8 | 0.4 |
| 4 | 18.5 | 1.4 | 56.5 | 0.4 | 7.8 | 1.7 | 8.6 | 0.4 |
| 6 | 19.1 | 1.4 | 40.6 | 0.4 | 6.6 | 1.6 | 6.6 | 0.4 |
| 8 | 17.8 | 1.4 | 31.2 | 0.4 | 5.7 | 1.5 | 5.3 | 0.4 |
| 10 | 16.0 | 1.4 | 25.7 | 0.4 | 5.0 | 1.5 | 4.5 | 0.4 |

Effect of encoding actor (device or host) on filtering throughput of single-GPU GateKeeper-GPU with increasing error threshold. The values represent the number of filtrations in terms of millions per second, produced by GateKeeper-GPU with respect to kernel or filter time. $e$ = error threshold.

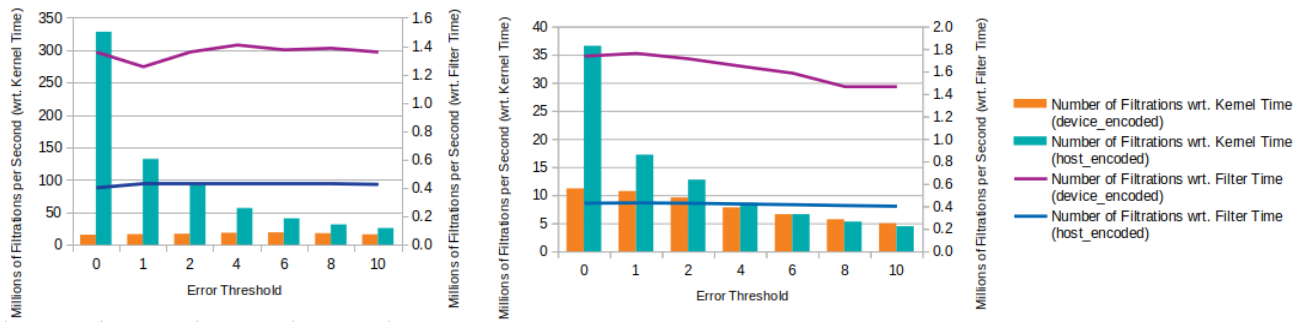

**Figure S. 14:** Effect of encoding actor (device or host) on filtering throughput of single-GPU GateKeeper-GPU with increasing error threshold for read length 250bp in Setup_1 and Setup_2, respectively

### C. Effect of Sequence Length on Filtering Throughput

Table S.20 represents the data depicted by Figure 7 for evaluating the effect of different sequence lengths on the performance of GateKeeper-GPU.

**Table S. 20:** Effect of Read Length on Filtering Throughput for Figure 7

| $e$ | Read Length | **Setup_1** | | **Setup_2** | |
|---|---|---|---|---|---|
| | | *Device-encoded* | *Host-encoded* | *Device-encoded* | *Host-encoded* |
| 0 | 100bp | 3.16 | 1.18 | 2.73 | 1.18 |
| | 150bp | 2.14 | 0.64 | 1.74 | 0.71 |
| | 250bp | 1.36 | 0.41 | 1.74 | 0.43 |
| 4 | 100bp | 3.16 | 1.23 | 2.43 | 1.11 |
| | 150bp | 2.18 | 0.68 | 1.65 | 0.70 |
| | 250bp | 1.41 | 0.43 | 1.65 | 0.43 |

Effect of read length on single-GPU GateKeeper-GPU's filtering throughput in terms of millions of filtrations per second. The values were calculated with respect to filter time. $e$ = error threshold.

### D. Multi-GPU Filtering Throughput Analysis

In order to see how GateKeeper-GPU's performance scales with increasing the number of GPGPU devices, we perform tests with 8 GPUs in Setup_1 and report the results in Table S.21 to S.23.

**Table S. 21:** Multi-GPU Filtering Throughput Analysis on 100bp for Figure 8

| wrt. | **Kernel Time** | | **Filter Time** | |
|---|---|---|---|---|
| **Number of GPU Devices** | **Device-encoded** | **Host-encoded** | **Device-encoded** | **Host-encoded** |
| **1** | 102 | 199 | 3 | 1 |
| **2** | 201 | 388 | 6 | 2 |
| **3** | 300 | 542 | 8 | 3 |
| **4** | 364 | 704 | 10 | 4 |
| **5** | 376 | 877 | 12 | 5 |
| **6** | 488 | 1,062 | 14 | 5 |
| **7** | 487 | 1,171 | 15 | 6 |
| **8** | 496 | 1,333 | 16 | 6 |

Multi-GPU filtering throughput of GateKeeper-GPU in terms of millions of filtrations per second, with respect to filter time and kernel time, in Setup_1. The error threshold is 2.

**Table S. 22:** Multi-GPU Filtering Throughput Analysis on 150bp for Figure 10 (b)

| wrt. | **Kernel Time** | | **Filter Time** | |
|---|---|---|---|---|
| **Number of GPU Devices** | **Device-encoded** | **Host-encoded** | **Device-encoded** | **Host-encoded** |
| **1** | 37 | 85 | 2 | 1 |
| **2** | 73 | 165 | 4 | 1 |
| **3** | 109 | 229 | 5 | 2 |
| **4** | 144 | 284 | 7 | 2 |
| **5** | 166 | 349 | 8 | 3 |
| **6** | 196 | 408 | 10 | 3 |
| **7** | 211 | 437 | 10 | 4 |
| **8** | 207 | 426 | 11 | 4 |

Multi-GPU filtering throughput of GateKeeper-GPU in terms of millions of filtrations per second, with respect to filter time and kernel time, in Setup_1. The error threshold is 4.

**Table S. 23:** Multi-GPU Filtering Throughput Analysis on 250bp for Figure 10 (c)

| Number of GPU Devices | **Kernel Time** | | **Filter Time** | |
|:---:|:---:|:---:|:---:|:---:|
| | **Device-encoded** | **Host-encoded** | **Device-encoded** | **Host-encoded** |
| **1** | 18 | 31 | 1 | 0 |
| **2** | 35 | 50 | 2 | 1 |
| **3** | 52 | 74 | 4 | 1 |
| **4** | 69 | 94 | 4 | 2 |
| **5** | 81 | 118 | 5 | 2 |
| **6** | 93 | 97 | 6 | 2 |
| **7** | 101 | 139 | 7 | 2 |
| **8** | 101 | 152 | 7 | 3 |

Multi-GPU filtering throughput of GateKeeper-GPU in terms of millions of filtrations per second, with respect to filter time and kernel time, in Setup_1. The error threshold is 8.

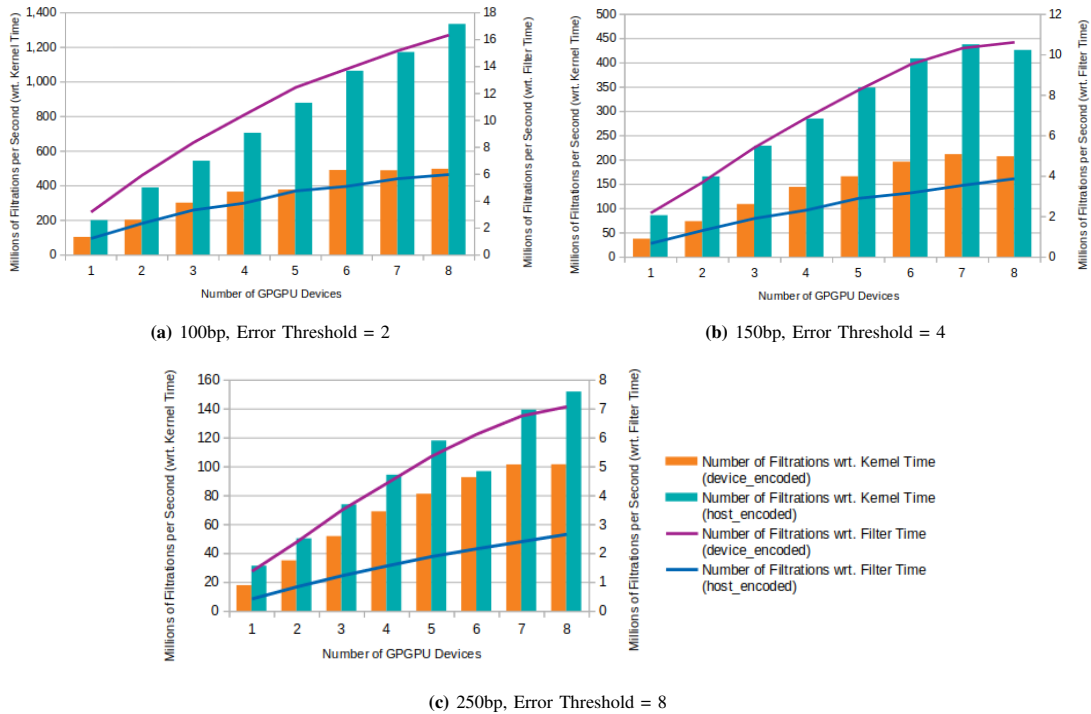

**(a)** 100bp, Error Threshold = 2

**(b)** 150bp, Error Threshold = 4

**(c)** 250bp, Error Threshold = 8

**Figure S. 15:** Multi-GPU filtering throughput of GateKeeper-GPU in Setup_1

## VII. Whole Genome Accuracy & Performance Analysis

**Table S. 24:** Whole genome mapping information with pre-alignment filtering on sim_set_1 (300bp)

*Mapping Information*

| mrFAST w/ | Mappings | Mapped Reads | Verification Pairs | Rejected Pairs (Reduction) |
|---|---|---|---|---|
| No Filter | 670 | 626 | 365,478,108 | NA |
| GateKeeper-GPU | 670 | 626 | 10,305,218 | 355,172,890 ( 97% ) |

*Time*

| mrFAST w/ | Filtering + DP Time / Speedup | | Overall Time / Speedup | |
|---|---|---|---|---|
| | *Setup_1* | *Setup_2* | *Setup_1* | *Setup_2* |
| No Filter | 0.04h | 0.05h | 0.12h | 0.12h |
| GateKeeper-GPU (d) | 0.13h / - | 0.12h / - | 0.31h / - | 1.10h / - |
| GateKeeper-GPU (h) | 0.12h / - | 0.12h / - | 0.31h / - | 1.06h / - |
| GateKeeper-FPGA | *279× | | *11× | |

Mapping information and time results by running mrFAST on the sim_set_1 with error threshold $e = 15$. Mapping information entries represent the number of corresponding metric. Speedup comparisons show mrFAST's performance by filtering sequence pairs using GateKeeper-GPU with single GPU (encoding in d: device, h: host) and GateKeeper-FPGA. DP: verification. *values were retrieved from GateKeeper [1] manuscript.

**Table S. 25:** Whole genome mapping information with pre-alignment filtering on sim_set_2 (150bp)

*Mapping Information*

| mrFAST w/ | Mappings | Mapped Reads | Verification Pairs | Rejected Pairs (Reduction) |
|---|---|---|---|---|
| No Filter | 34,677,103 | 918,403 | 10,379,001,396 | NA |
| GateKeeper-GPU | 34,677,011 | 918,403 | 962,733,131 | 9,416,268,265 ( 90% ) |

*Time*

| mrFAST w/ | Filtering + DP Time / Speedup | | Overall Time / Speedup | |
|---|---|---|---|---|
| | *Setup_1* | *Setup_2* | *Setup_1* | *Setup_2* |
| No Filter | 1.15h | 1.17h | 2.07h | 2.13h |
| GateKeeper-GPU (d) | 0.38h / 3.0× | 0.97h / 1.2× | 1.96h / 1.1× | 2.23h / - |
| GateKeeper-GPU (h) | 0.33h / 3.4× | 0.93h / 1.2× | 1.49h / 1.4× | 2.19h / - |

Mapping information and time results by running mrFAST on the sim_set_2 with error threshold $e = 8$. Mapping information entries represent the number of corresponding metric for mapping information for mapping information. Speedup comparisons show mrFAST's performance by filtering sequence pairs using GateKeeper-GPU with single GPU (encoding in d: device, h: host).

## VIII. Resource Utilization and Power Analysis

**Table S. 26:** Power Consumption of GateKeeper-GPU in Setup_2

| Power (mW) | Device-encoded | | Host-encoded | |
|---|---|---|---|---|
| | *100bp* | *250bp* | *100bp* | *250bp* |
| min | 30,193 | 30,097 | 30,194 | 30,097 |
| max | 107,756 | 123,026 | 125,849 | 129,799 |
| average | 77,699 | 85,532 | 74,713 | 77,684 |

Power Consumption (milliwatts) for single GPGPU in Setup_2. The values were obtained by running CUDA command-line profiler nvprof.



\end{document}